\documentclass[twocolumn,superscriptaddress]{revtex4-1}
\usepackage{float}
\usepackage{amsmath}
\usepackage{amssymb}
\usepackage{mathtools}
\usepackage{amsfonts}
\usepackage{amsthm}
\usepackage{bm}
\usepackage{braket}
\usepackage{cases}
\usepackage{comment}
\usepackage{graphicx,color}
\usepackage{subfigure}
\usepackage{enumitem}
\usepackage{color}

\def\({\left(}

\def\){\right)}

\begin{document}
\title{Finite-size security of continuous-variable quantum key distribution with digital signal processing}

\author{Takaya Matsuura}
\affiliation{Department of Applied Physics, Graduate School of Engineering,
The University of Tokyo, 7-3-1 Hongo, Bunkyo-ku, Tokyo 113-8656, Japan}

\author{Kento Maeda}
\affiliation{Department of Applied Physics, Graduate School of Engineering,
The University of Tokyo, 7-3-1 Hongo, Bunkyo-ku, Tokyo 113-8656, Japan}

\author{Toshihiko Sasaki}
\affiliation{Department of Applied Physics, Graduate School of Engineering,
The University of Tokyo, 7-3-1 Hongo, Bunkyo-ku, Tokyo 113-8656, Japan}
\affiliation{Photon Science Center, Graduate School of Engineering, The University of Tokyo, 7-3-1 Hongo, Bunkyo-ku, Tokyo 113-8656, Japan}

\author{Masato Koashi}
\affiliation{Department of Applied Physics, Graduate School of Engineering,
The University of Tokyo, 7-3-1 Hongo, Bunkyo-ku, Tokyo 113-8656, Japan}
\affiliation{Photon Science Center, Graduate School of Engineering, The University of Tokyo, 7-3-1 Hongo, Bunkyo-ku, Tokyo 113-8656, Japan}

\begin{abstract}
  In comparison to conventional discrete-variable (DV) quantum key distribution (QKD), continuous-variable (CV) QKD with homodyne/heterodyne measurements has distinct advantages of lower-cost implementation and affinity to wavelength division multiplexing. On the other hand, its continuous nature makes it harder to accommodate to practical signal processing, which is always discretized, leading to lack of complete security proofs so far. Here we propose a tight and robust method of estimating fidelity of an optical pulse to a coherent state via heterodyne measurements. We then construct a binary phase modulated CV QKD protocol and prove its security in the finite-key-size regime against general coherent attacks, based on proof techniques of DV QKD. Such a complete security proof achieves a significant milestone in exploiting the benefits of CV QKD.
\end{abstract}


\maketitle

Quantum key distribution (QKD) aims at generating a secret key shared between two remote legitimate parties with information-theoretic security, which provides secure communication against an adversary with arbitrary computational power and hardware technology.  Since the first proposal in 1984 \cite{BB84}, various QKD protocols have been proposed with many kinds of encoding and decoding schemes. 
These protocols are typically classified into two categories depending on the detection methods. One of them is called discrete-variable (DV) QKD, which uses photon detectors and includes earlier protocols such as BB84 \cite{BB84} and B92 \cite{B92} protocols. The other is called continuous-variable (CV) QKD, which uses homodyne and heterodyne measurements with photo detectors \cite{Ralph1999,Hillery2000,GG02}. 
Although DV QKD is more mature and achieves a longer distance if photon detectors with low dark count rates are available, CV QKD has its own distinct advantages for a short distance.
It can be implemented with components common to coherent optical communication technology and is expected to be cost-effective. Excellent spectral filtering capability inherent in homodyne/heterodyne measurements suppresses crosstalk in wavelength division multiplexing (WDM) channels.
This allows multiplexing of hundreds of QKD channels into a single optical fiber \cite{Eriksson2020} as well as co-propagation with classical data channels \cite{Huang2015,Kumar2015,Huang2016,Karinou2017,Karinou2018,Eriksson2018,Eriksson2019}, which makes integration into existing communication network easier.

One major obstacle in putting CV QKD to practical use is the gap between the employed continuous variables and mandatory digital signal processing. The CV QKD protocols are divided into two branches depending on whether the modulation method of the encoder is also continuous, or it is discrete. The continuous modulation protocols usually adopts Gaussian modulation, in which the sender chooses the complex amplitude of a coherent-state pulse according to a Gaussian distribution \cite{Ralph1999,Hillery2000,GG02,Grosshans2003,hetero04} (see Ref.~\cite{CV_protocol_review} for a review).  This allows powerful theoretical tools such as Gaussian optimality \cite{CV_gaussian_optimality1,CV_gaussian_optimality2}, and complete security proofs for a finite-size key and against general attacks have been given \cite{Gaussian_unitary}.  To implement Gaussian protocols with a digital random-number generator and digital signal processing, it is necessary to approximate the continuous distribution with a constellation composed of a large but finite number of complex amplitudes \cite{Lev,Kaur2019}. This is where difficulty arises, and the security analysis has been confined to the asymptotic regime and collective attacks. The other branch gives priority to simplicity of the modulation and uses a very small (usually two to four) number of amplitudes \cite{Silberhorn2002,Hirano2003,Leverrier2009}. As for the security analysis, the status is more or less similar to the Gaussian constellation case, and current security proofs are either in the asymptotic regime against collective attacks \cite{two_state_Lutken,three_state,four_state_Lutken,four_state_Leverrier} or in the finite-size regime but against more restrictive attacks \cite{Papanastasiou2019}. Hence, regardless of approaches, a complete security proof of CV QKD in the finite-size regime against general attacks has been a significant milestone yet to be achieved.

Here we mark the above milestone by proposing a binary phase-modulated CV QKD protocol with a complete security proof in the finite-size regime against general attacks. The key ingredient is a novel estimation method using heterodyne measurements which is suited for analysis of confidence region in the finite-size regime. The outcome of heterodyne measurement, which is unbounded, is converted to a bounded value by a smooth function such that its expectation is proved to be no larger than the fidelity of the input pulse to a coherent state. This allows us to use a standard technique to derive a lower bound on the fidelity with a required confidence level in the finite-size regime. The fidelity as a measure of disturbance in the binary modulated protocol is essentially the same as what is monitored through bit errors in the B92 protocol \cite{B92,Tamaki2003,Koashi2004}. This allows us to construct a security proof based on a reduction to distillation of entangled qubit pairs \cite{Shor_Preskill,Lo1999}, which is a technique frequently used for DV QKD protocols.

\newpage

\noindent{\bf \large Results}\\
{\bf Estimation of fidelity to a coherent state.}
We first introduce a test scheme to estimate the fidelity between an input optical state $\rho$ and the vacuum state $\ket{0}\!\bra{0}$ through a heterodyne measurement. For an input state $\rho$ of a single optical mode, the heterodyne measurement produces an outcome $\hat{\omega} \in \mathbb{C}$ with a probability density
\begin{equation}
    q_{\rho}(\omega)\, d^2\omega\coloneqq \bra{\omega}\rho\ket{\omega} \frac{d^2\omega}{\pi},
    \label{eq:hetero_density}
\end{equation}
where a coherent state $\ket{\omega}$ is defined as 
\begin{equation}
    \label{eq:definition_coherent_state}
    \ket{\omega}\coloneqq e^{-|\omega|^2/2}\sum_{n=0}^{\infty}\frac{\omega^n}{\sqrt{n!}}\ket{n}.
\end{equation}
We denote the expectation associated with the distribution $q_{\rho}(\omega)$ simply by $\mathbb{E}_\rho$.
To construct a lower bound for the fidelity $\bra{0}\rho\ket{0}$ from $\hat{\omega}$, we will use the associated Laguerre polynomials which are given by
\begin{equation}
  \label{eq:associate_Laguerre}
  L_n^{(k)}(\mu)\coloneqq (-1)^k\frac{d^kL_{n+k}(\mu)}{d\mu^k},
\end{equation}
where
\begin{equation}
    L_n(\mu)\coloneqq \frac{e^\mu}{n!}\frac{d^n}{d\mu^n}(e^{-\mu}\mu^n)
\end{equation}
are the Laguerre polynomials.
Our test scheme is based on the following theorem.

\medskip

\noindent{\it Theorem 1:} Let $\Lambda_{m,r}(\mu)$ be a bounded function given by 
\begin{equation}
    \label{eq:postprocess_function}
  \Lambda_{m,r}(\mu)\coloneqq e^{-r\mu}(1+r)L_m^{(1)}((1+r)\mu),
\end{equation}
for an integer $m\geq 0$ and a real number $r>0$. 
Then, we have
\begin{equation}
  \label{eq:theorem1_estimation}
  \mathbb{E}_\rho[\Lambda_{m,r}(|\hat{\omega}|^2)]=\bra{0}\rho\ket{0}+\sum_{n=m+1}^{\infty}\frac{\bra{n}\rho\ket{n}}{(1+r)^n}I_{n,m},
\end{equation}
where $I_{n,m}$ are constants satisfying $(-1)^m I_{n,m}>0$.

\medskip

\begin{figure}
  \centering
    \includegraphics[width=0.95\linewidth]{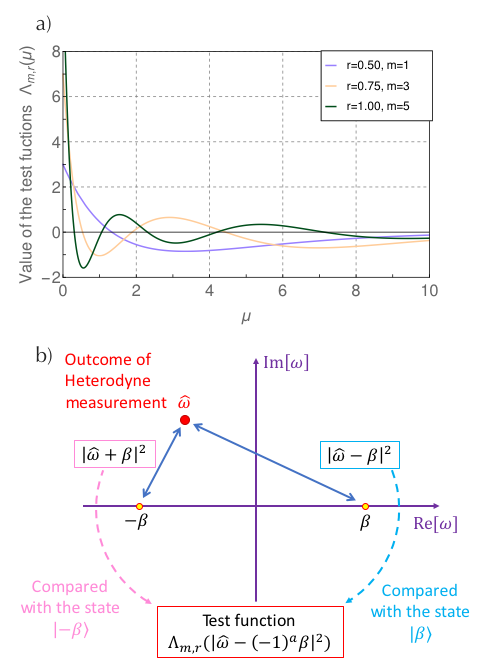}
    \caption{Illustration of the test scheme to estimate the fidelity. {\rm a) Example of the test functions used in the estimation.  In general, the range of the function $\Lambda_{m,r}$ gets larger when $m$ gets larger.  b) A schematic description of the usage of obtained outcomes in heterodyne measurement. In order to estimate the lower bound on the fidelity to the coherent states $\ket{\pm \beta}$, the squared distance between the outcome $\hat{\omega}$ and the objective point $(-1)^{a}\beta$ (i.e., $|\hat{\omega} - (-1)^{a}\beta|^2$) is used.} }
   \label{fig:test_scheme}
\end{figure}  

From Eq.~\eqref{eq:theorem1_estimation}, a lower bound on the fidelity between $\rho$ and the vacuum state is given by
\begin{equation}
  \label{eq:expectation_test_function}
  \mathbb{E}_{\rho}[\Lambda_{m,r}(|\hat{\omega}|^2)]\leq \bra{0}\rho\ket{0}\quad (m:\ {\rm odd})
\end{equation} 
for any odd integer $m$. 
As seen in Figure~\ref{fig:test_scheme}.~a), the absolute value and the slope of the function $\Lambda_{m,r}$ are moderate for small values of $m$ and $r$, which is advantageous in executing the test in a finite duration with a finite resolution.  Compared to a similar method proposed in \cite{fidelity_prev}, our method excels in its tightness for weak input signals; we see from Eq.~\eqref{eq:theorem1_estimation} that, regardless of the value of $r$, the inequality \eqref{eq:expectation_test_function} saturates when $\rho$ has at most $m$ photons.  This is crucial for the use in QKD in which tightness directly affects the efficiency of the key generation.  

Extension to the fidelity to a coherent state $\ket{\beta}$ is straightforward as
\begin{equation}
  \mathbb{E}_{\rho}[\Lambda_{m,r}(|\hat{\omega}-\beta|^2)]\leq {\rm Tr}\(\rho\ket{\beta}\!\bra{\beta}\) \quad (m:\ {\rm odd}).
  \label{eq:distance_coherent}
\end{equation}
The proofs are given in Methods.

\begin{figure}
  \centering
    \includegraphics[width=0.95\linewidth]{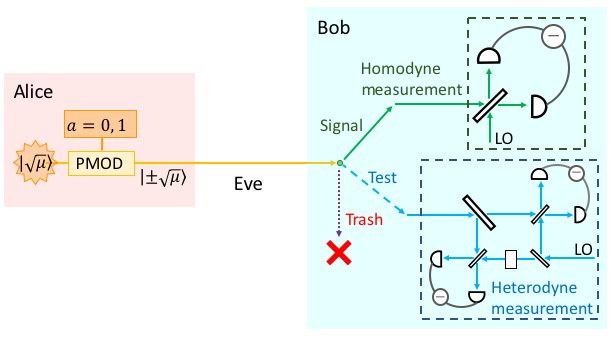}
    \caption{Illustration of the proposed continuous-variable QKD protocol. {\rm Alice generates a random bit $a\in\{0,1\}$ and sends a coherent state with amplitude $(-1)^a \sqrt{\mu}$. Bob chooses one of the three measurements based on the predetermined probability. In the signal round , Bob performs a homodyne measurement on the received optical pulse and obtains an outcome $\hat{x}$.  In the test round, Bob performs a heterodyne measurement on the received optical pulse and obtains an outcome $\hat{\omega}$.  In the trash round, he produces no outcome.}}
   \label{setup}
\end{figure}

\noindent{\bf Proposed protocol.}
Based on this fidelity test, we propose the following discrete-modulated protocol (see Figure~\ref{setup}). In what follows, Alice and Bob predetermine the number of rounds $N$, the protocol parameters $(\mu, p_{\rm sig}, p_{\rm test}, p_{\rm trash}, \beta, s)$, the acceptance probability of homodyne measurement $f_{\rm suc}(|x|)\ (x \in \mathbb{R})$ with $f_{\rm suc}(0)=0$, and the parameters for the test function $(m, r)$.  We assume all the parameters are positive and that $p_{\rm sig}+p_{\rm test}+p_{\rm trash}=1$.
\begin{enumerate}
  \item Alice generates a random bit $a\in\{0,1\}$ and sends an optical pulse $B$ in a coherent state with amplitude $(-1)^a \sqrt{\mu}$ to Bob.  She repeats it $N$ times.
  \item For each of the received $N$ pulses, Bob chooses a label from $\rm \{signal, test, trash\}$ with probabilities $p_{\rm sig},p_{\rm test}$, and $p_{\rm trash}$, respectively. According to the label, Alice and Bob do one of the following procedures.
  \begin{itemize}
    \item[[signal\textrm{]}] Bob performs a homodyne measurement on the received optical pulse, and obtains an outcome $\hat{x} \in \mathbb{R}$.  With a probability $f_{\rm suc}(|\hat{x}|)$, he regards the detection to be a ``success'', and defines a bit $b=0$ (resp.~$1$) when ${\rm sign}(\hat{x})=+(-)1$.  He announces success/failure of the detection.  In the case of a success, Alice (resp.~Bob) keeps $a$ ($b$) as a sifted key bit.
    \item[[test\textrm{]}] Bob performs a heterodyne measurement on the received optical pulse, and obtains an outcome $\hat{\omega}$.  Alice announces her bit $a$.  Bob calculates the value of $\Lambda_{m,r}(|\hat{\omega} -(-1)^a\beta|^2)$.
    \item[[trash\textrm{]}] Alice and Bob produce no outcomes.
  \end{itemize}
  \item We denote the numbers of ``success'' and ``failure'' signal rounds, test rounds, and trash rounds by $\hat{N}^{\rm suc}, \hat{N}^{\rm fail}, \hat{N}^{\rm test}$, and $\hat{N}^{\rm trash}$, respectively.  ($N= \hat{N}^{\rm suc}+\hat{N}^{\rm fail}+\hat{N}^{\rm test}+\hat{N}^{\rm trash}$ holds by definition.) 
  Bob calculates the sum of $\Lambda_{m,r}(|\hat{\omega} -(-1)^a\beta|^2)$ obtained in the $\hat{N}^{\rm test}$ test rounds, which we denote by $\hat{F}$.
  \item For error correction, they use $(H_{\rm EC}+s')$ bits of encrypted communication consuming a pre-shared secret key to do the following.  Alice sends Bob $H_{\rm EC}$ bits of syndrome of a linear code for her sifted key.  Bob reconciles his sifted key accordingly. Alice and Bob verify the correction by comparing $s'$ bits via universal$_2$ hashing \cite{univhash}.
  \item 
  Bob computes and announces the final key length by
  \begin{equation}
    \qquad \ \, \hat{N}^{\rm fin} = \hat{N}^{\rm suc} \(1 - h\(U(\hat{F}, \hat{N}^{\rm trash})/\hat{N}^{\rm suc}\) \) - s,
    \label{eq:final_key_length}
  \end{equation}
  where $h(x)\coloneqq -x {\rm log}_2(x)-(1-x){\rm log}_2(1-x)$ is the binary entropy function and the function $U(\hat{F}, \hat{N}^{\rm trash})$ will be specified below.  Alice and Bob apply privacy amplification to obtain the final key.  The net key gain $\hat{G}$ per pulse is therefore given by
  \begin{equation}
    \hat{G} = (\hat{N}^{\rm fin} - H_{\rm EC} - s') / N.
  \end{equation}
\end{enumerate}

\begin{figure}
  \begin{center}
    \includegraphics[width=0.95\linewidth]{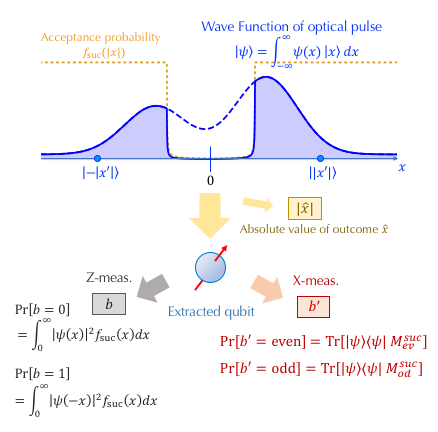}
    \caption{Illustration of Bob's qubit extraction in the virtual procedure.  {\rm Bob performs on the optical pulse a non-demolition projective measurement, with which the absolute value of the outcome of homodyne measurement $|\hat{x}|$ is determined. Then, Bob extracts a qubit $B$ by the operation $\mathcal{F}$ defined in Eq.~\eqref{eq:F}. A $Z$-basis measurement on this qubit gives the same sifted key bit $b$ as described in the original protocol. On the other hand, the $X$-basis measurement on this qubit reveals the parity of photon number of the received optical pulse.}}
   \label{fig:qubit_extraction}
  \end{center}
\end{figure}

\noindent{\bf Security Proof.}
We determine a sufficient amount of the privacy amplification according to Shor and Preskill [17], which has been widely used for the DV-QKD protocols. We consider an equivalent protocol in which Alice and Bob determine their sifted key bits $a$ and $b$ by measurement on a pair of qubits. 
For Alice, we introduce a qubit $A$ and assume that she entangles it with an optical pulse $B$ in a state
\begin{equation}
  \ket{\Psi}_{AB}\coloneqq \frac{\ket{0}_A\ket{\sqrt{\mu}}_{B}+\ket{1}_A\ket{-\sqrt{\mu}}_{B}}{\sqrt{2}}.
  \label{eq:prepared_state}
\end{equation}
Then, Step~1.\ is equivalent to the preparation of $\ket{\Psi}_{AB}$ followed by a measurement of the qubit $A$ on $Z$ basis $\{\ket{0},\ket{1}\}$ to determine the bit value $a$.
For Bob, we construct a process of probabilistically converting the received optical pulse $B$ to a qubit $B'$ (See Figure~\ref{fig:qubit_extraction}).  Consider a completely positive map defined by 
\begin{equation}
  \label{eq:F}
  \mathcal{F}_{B\to B'}(\rho_{B})\coloneqq \int_{0}^{\infty} dx\, K^{(x)}\rho_B K^{(x)\dagger},
\end{equation}
with
\begin{equation}
    \label{eq:K_x}
    K^{(x)}\coloneqq \sqrt{f_{\rm suc}(x)}\left(\ket{0}_{B'}\!\bra{x}_{B}+\ket{1}_{B'}\!\bra{-x}_{B}\right).
\end{equation}
When the pulse $B$ is in a state $\rho_B$, the corresponding process succeeds with a probability $p_{\rm suc}$ and then prepares the qubit $B'$ in a state $\rho_{B'}$, where $p_{\rm suc} \rho_{B'} = \mathcal{F}_{B\to B'}(\rho_B)$. If the qubit B’ is further measured on $Z$ basis, probabilities of the outcome $b=0,1$ are given by
\begin{align}
  p_{\rm suc}\bra{0} \rho_{B'} \ket{0} &= \int_{0}^{\infty} f_{\rm suc}(x) dx\, \bra{x}\rho_B\ket{x},\label{eq:success_b_0}\\
  p_{\rm suc}\bra{1} \rho_{B'} \ket{1} &= \int_{0}^{\infty} f_{\rm suc}(x) dx\, \bra{-x}\rho_B\ket{-x},\label{eq:success_b_1}
\end{align}
which shows the equivalence to the signal round in Step~2.
This is illustrated in Figure~\ref{fig:qubit_extraction}.

Once the qubit pair $AB'$ are introduced, the amount of privacy amplification is connected to the so-called phase error rate. Instead of $Z$-basis measurements in the equivalent protocol, consider a virtual protocol in which the qubits are measured on $X$ basis $\{\ket{\pm}\coloneqq (\ket{0}+\ket{1})/\sqrt{2} \}$.  A pair with outcomes $(+,-)$ or $(-,+)$ is defined to be a phase error.  Let $\hat{N}^{\rm suc}_{\rm ph}$ be the number of phase errors among $\hat{N}^{\rm suc}$ pairs.
If we have a good upper bound $e_{\rm ph}$ on the phase error rate $\hat{N}^{\rm suc}_{\rm ph}/ \hat{N}^{\rm suc}$, shortening by fraction $h(e_{\rm ph})$ via privacy amplification achieves the security in the asymptotic limit \cite{Shor_Preskill,complementarity}.
To cover the finite-size cases as well, our goal is to construct $U(\hat{F}, \hat{N}^{\rm trash})$ which satisfies
\begin{equation}
  {\rm Pr}\left[\hat{N}_{\rm ph}^{\rm suc}\leq U(\hat{F}, \hat{N}^{\rm trash})\right]\geq 1-\epsilon \label{eq:probability_condition}
\end{equation}
for any attack in the virtual protocol. 
It is known that it immediately implies that the actual protocol can be made 
$\epsilon_\mathrm{sec}$-secure with a small security parameter
$\epsilon_\mathrm{sec}=\sqrt{2}\sqrt{\epsilon+2^{-s}}+2^{- s'}$\cite{complementarity,Hayashi_Tsurumaru}.
See Methods for the detailed definition of security.

At this point, it is beneficial for the analysis of the phase error statistics to clarify what property of the optical pulse $B$ is measured by Bob's $X$-basis measurement (see Figure~\ref{fig:qubit_extraction}).  Let $\Pi_{\rm ev(od)}$ be the projection to the subspace with even (resp.\ odd) photon numbers.  ($\Pi_{\rm ev}+\Pi_{\rm od}=\bm{1}_B$ holds by definition.)  Furthermore, since $\Pi_{\rm ev} - \Pi_{\rm odd}$ is the operator for an optical phase shift of $\pi$, we have $(\Pi_{\rm ev} - \Pi_{\rm odd})\ket{x}=\ket{-x}$.  Eq.~\eqref{eq:K_x} is then rewritten as
\begin{equation}
  \label{eq:K_x_xbasis}
  K^{(x)}= \sqrt{2 f_{\rm suc}(x)}\left(\ket{+}_{B'}\!\bra{x}_{B}\Pi_{\rm ev}+\ket{-}_{B'}\!\bra{x}_{B}\Pi_{\rm od}\right).
\end{equation}
Therefore, the probability of obtaining $+(-)$ in the $X$-basis measurement is given by 
\begin{equation}
    \bra{+(-)}\mathcal{F}_{B\to B'}(\rho_{B})\ket{+(-)} = {\rm Tr}\(\rho_B M^{\rm suc}_{{\rm ev(od)}}\),
\end{equation}
where
\begin{equation}
      M_{{\rm ev(od)}}^{\rm suc}\coloneqq \int_{0}^{\infty} 2 f_{\rm suc}(x)dx\, \Pi_{\rm ev(od)}\ket{x}_{B}\!\bra{x}\Pi_{\rm ev(od)}.
      \label{eq:parity_measurement}
\end{equation}
This shows that Bob's $X$-basis measurement distinguishes the parity of the photon number of the received pulse.
In this sense, the secrecy of our protocol is assured by the complementarity between the sign of the quadrature and the parity of the photon number.

For the construction of $U(\hat{F}, \hat{N}^{\rm trash})$, we consider a modified scenario as follows.
\begin{enumerate}
  \item[1'.] Alice prepares a qubit $A$ and an optical pulse $B$ in a state $\ket{\Psi}_{AB}$ defined in \eqref{eq:prepared_state}.  She repeats it $N$ times.
  \item[2'.] According to the label announced by Bob in the same way as in Step~2., Alice and Bob do one of the following procedures.
  \begin{itemize}
    \item[[signal\textrm{]}] Bob makes a measurement on the received pulse $B$ specified by measurement operators $\{M^{\rm suc}_{\rm ev}, M^{\rm suc}_{\rm od}, 1- M^{\rm suc}_{\rm ev}- M^{\rm suc}_{\rm od}\}$ to determine success/failure of detection, and the parity $b'=\text{even/odd}$ upon success.  He announces success/failure of detection. Alice measures her qubit $A$ on $X$ basis to obtain $a'=+/-$.
    \item[[test\textrm{]}] Bob performs a heterodyne measurement on the received optical pulse, and obtains an outcome $\hat{\omega}$.  Alice announces her bit $a$.  Bob calculates the value of $\Lambda_{m,r}(|\hat{\omega} -(-1)^a\beta|^2)$.
    \item[[trash\textrm{]}] Alice measures her qubit $A$ on $X$ basis to obtain $a'=+/-$.
  \end{itemize}
  \item[3'.] $\hat{N}^{\rm suc}, \hat{N}^{\rm trash}$, and $\hat{F}$ are defined as in Step~3.  Let $\hat{N}^{\rm suc}_{\rm ph}$ be the number of rounds in the $\hat{N}^{\rm suc}$ success rounds with $(a', b')=(+, {\rm odd})$ or $(-, {\rm even})$.
  Let $\hat{Q}_{-}$ be the number of rounds in the $\hat{N}^{\rm trash}$ trash rounds with $a'=-$.
\end{enumerate}
When the adversary adopts the same attack strategy on the virtual protocol and the modified scenario, the marginal joint probability of $(\hat{N}^{\rm suc}_{\rm ph}, \hat{F}, \hat{N}^{\rm trash})$ should be the same.  Hence it suffices to prove Eq.~\eqref{eq:probability_condition} for the modified scenario.  

In order to bound $\hat{N}^{\rm suc}_{\rm ph}$, we seek an upper bound on a linear combination of variables, 
\begin{equation}
\hat{T}[\kappa, \gamma]:= p_{\rm sig}^{-1} \hat{N}^{\rm suc}_{\rm ph} + p_{\rm test}^{-1} \kappa \hat{F} - p_{\rm trash}^{-1} \gamma \hat{Q}_{-}
\label{eq:def_T}
\end{equation}
with coefficients $\kappa, \gamma\ge 0$ which are independent of the observed values of $\hat{F}$ and $\hat{N}^{\rm trash}$.
First, the expectation $\mathbb{E}\bigl[\hat{T}[\kappa, \gamma]\bigr]$ can be bounded as follows.
Let $\rho_{AB}$ be the state of the qubit $A$ and the received pulse $B$ averaged over $N$ pairs, and define relevant operators as
\begin{align}
  &M_{\rm ph}^{\rm suc}\coloneqq \ket{+}\!\bra{+}_A\otimes M_{{\rm od}}^{\rm suc} + \ket{-}\!\bra{-}_A\otimes M_{{\rm ev}}^{\rm suc},\label{eq:m_ph}\\
  &\Pi_{\rm fid}\coloneqq \ket{0}\!\bra{0}_{A}\otimes\ket{\beta}\!\bra{\beta}_{B}+\ket{1}\!\bra{1}_{A}\otimes\ket{-\beta}\!\bra{-\beta}_{B},\label{eq:pi_fid} \\
  &\Pi_{\pm}\coloneqq \ket{\pm}\!\bra{\pm}_A\otimes \bm{1}_B, \label{eq:pi_pm}
\end{align}
and
\begin{equation}
  M[\kappa, \gamma]:=M^{\rm suc}_{\rm ph} +\kappa \Pi_{\rm fid} - \gamma \Pi_{-}. \label{eq:def_of_M}
\end{equation}
Then we immediately have 
\begin{equation}
  \mathbb{E}[ \hat{N}^{\rm suc}_{\rm ph} ]= p_{\rm sig} N\, \mathrm{Tr}\(\rho_{AB} M^{\rm suc}_{\rm ph}\) \label{eq:expectation_n_ph}
\end{equation}
and 
\begin{equation}
\mathbb{E}[\hat{Q}_{-}]= p_{\rm trash} N\, \mathrm{Tr} \(\rho_{AB}\Pi_{-} \), \label{eq:expectation_q}
\end{equation}
while application of the property of Eq.~\eqref{eq:distance_coherent} leads to
\begin{equation}
  \mathbb{E}[\hat{F}] \leq  p_{\rm test} N \,\mathrm{Tr} \(\rho_{AB}\Pi_{\rm fid}\).\label{eq:expectation_f}
\end{equation}
Hence we have $\mathbb{E}\bigl[\hat{T}[\kappa, \gamma]\bigr]\leq N\, \mathrm{Tr} \(\rho_{AB} M[\kappa, \gamma]\)$.
If we can find a constant $B(\kappa, \gamma)\in \mathbb{R}$ satisfying the operator inequality 
\begin{equation}
  M[\kappa, \gamma]\leq B(\kappa, \gamma) \bm{1}_{AB}, \label{eq:operator_inequality}
\end{equation}
we obtain a bound $\mathbb{E}\bigl[\hat{T}[\kappa, \gamma]\bigr]\leq N B(\kappa, \gamma)$, which is indepenedent  of $\rho_{AB}$.
An easily computable bound $B(\kappa, \gamma)$ is derived in Methods. 
Then we expect that
\begin{equation}
  \hat{T}[\kappa, \gamma] \leq N B(\kappa, \gamma) + \delta_1(\epsilon/2)
  \label{eq:bound_for_T}
\end{equation}
holds with a probability no smaller than $1-\epsilon/2$.
Here, the term  $\delta_1(\epsilon/2)$ of $O(\sqrt{N})$ allows for fluctuations 
from finite-size effects, and is determined by using Azuma's inequality  \cite{Azuma_ineq} (see Methods).
Although Eq.~\eqref{eq:def_T} includes $\hat{Q}_{-}$ which is inaccessible in the actual protocol, we can derive a bound by noticing that it is an outcome from Alice's qubits 
and is independent of the adversary's attack.  In fact, given $\hat{N}^{\rm trash}$, 
it is the tally of $\hat{N}^{\rm trash}$ Bernoulli trials with a probability 
$\| \bra{-}_A \ket{\Psi}_{AB} \|^2 = (1-e^{-2\mu})/2 \eqqcolon q_{-}$. 
Hence, we can derive an inequality of the form 
\begin{equation}
\hat{Q}_- \leq  q_{-} \hat{N}^{\rm trash}  + \delta_2(\epsilon/2; \hat{N}^{\rm trash}) \label{eq:bound_Q}
\end{equation}
which holds with a probability no smaller than $1- \epsilon/2$.
Here $\delta_2(\epsilon/2; \hat{N}^{\rm trash})$ can be determined by a Chernoff bound (see Methods).
Combining Eqs.~\eqref{eq:def_T}, \eqref{eq:bound_for_T}, and \eqref{eq:bound_Q}, we obtain $U(\hat{F} , \hat{N}^{\rm trash} )$ satisfying Eq.~\eqref{eq:probability_condition} to complete the finite-size security proof.

\noindent{\bf Numerical Simulation.}
\begin{figure}
  \centering
  \includegraphics[height=0.838\textheight]{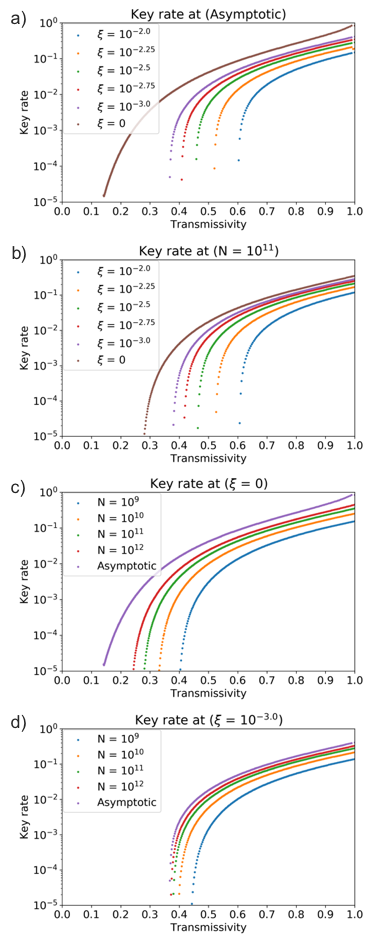}
  \caption{The net key gain per pulse $\hat{G}$ (key rate) as a function of transmissivity of the optical path. {\rm We assumed that the optical pulse which Bob receives is randomly displaced coherent state with a Gaussian distribution with the variance $\xi/2$. a) The asymptotic key rate for the different values of $\xi$. b) The key rate for the different values of $\xi$ when the pulse number is finite $(N =10^{11})$.  c) The key rate without the excess noise ($\xi = 0$).  d) The key rate with the excess noise of $\xi=10^{-3.0}$.}}
  \label{fig:numerical_simulation}
\end{figure}
We simulated the net key gain per pulse $\hat{G}$ as a function of transmissivity of the optical path $\eta$ (including the efficiency of Bob's apparatus).  We assume a channel model with a loss with transmissivity $\eta$ and an excess noise for Bob's apparatus with which the received state is displaced randomly to increase the variance by a factor of $1+\xi$.
We assume a step function with a threshold $x_{\rm th}(>0)$ as the acceptance probability $f_{\rm suc}(|x|)$. The expected amplitude of coherent state $\beta$ is chosen to be $\sqrt{\eta\mu}$.  
We set $\epsilon_{\rm sec} = 2^{-50}$ for the security parameter, and set $\epsilon=2^{-s}=\epsilon_{\rm sec}^2/16$ and $2^{-s'}=\epsilon_{\rm sec}/2$.  We thus have two coefficients $(\kappa, \gamma)$, four protocol parameters $(\mu, x_{\rm th}, p_{\rm sig}, p_{\rm test})$, and two parameters $(m,r)$ of the test function to be determined.  For each transmissivity $\eta$, we determined $(\kappa,\gamma)$ via a convex optimization using the CVXPY 1.0.25 \cite{cvxpy,cvxpy_rewriting} and $(\mu, x_{\rm th}, p_{\rm sig}, p_{\rm test})$ via the Nelder-Mead in the scipy.minimize library in Python, in order to maximize the key rate.  Furthermore, we adopted $m=1$ and $r=0.412019$, which leads to $({\rm max}\Lambda_{m,r},{\rm min}\Lambda_{m,r}) = (2.82404, -0.993162)$.
See Methods for the detail of the model of our numerical simulation.

Figure~\ref{fig:numerical_simulation} shows the key rates of our protocol in the asymptotic limit $N \to \infty$ and finite-size cases with $N =10^{9}\text{--}10^{12}$ for $\xi=10^{-2.0} \text{--} 10^{-3.0}$ and $0$.  
For the noiseless model ($\xi = 0$), the asymptotic rate reaches $\eta=0.2$. In the case of $\xi = 10^{-3.0}$, it reaches $\eta=0.4$, which is comparable to the result of a similar binary modulation protocol \cite{two_state_Lutken}. 
As for finite-size key rates, we see that the noiseless model shows a significant finite-size effect even for $N=10^{12}$.  On the other hand, with a presence of noises ($\xi=10^{-3.0}$) the effect becomes milder, and $N=10^{11}$ is enough to achieve a rate  close to the asymptotic case.

\medskip
\noindent{\bf \large Discussion.} 
Numerically simulated key rates above were computed on the implicit assumption that Bob's observed quantities are processed with infinite precision.  Even when these are approximated with a finite set of discrete points, we can still prove the security with minimal degradation of key rates.  For the heterodyne measurement used for the test in the protocol, assume that a digitized outcome $\omega_{\rm dig}$ ensures that the true value $\hat{\omega}$ lies in a range $\Omega(\omega_{\rm dig})$.  Then, we need only to replace $\Lambda_{m,r}(|\hat{\omega}\pm\beta|^2)$ with its worst-case value, $\min \{ \Lambda_{m,r}(|\hat{\omega} \pm \beta|^2) : \hat{\omega} \in \Omega (\omega_{\rm dig}) \}$.
As seen in Figure~\ref{fig:test_scheme}.~a), the slope of function $\Lambda_{m,r}(\mu)$ is moderate and goes to zero for $\mu \to \infty$. This means that the worst-case value can be made close to the true value, leading to small influence on the key rate.
For the homodyne measurement used for the signal, finite precision can be treated through appropriate modification of the acceptance probability $f_{\rm suc}(x)$.
Aside from a very small change in the success rate and the bit error rate, this function affects the key rate only through integrals in Eqs.~\eqref{eq:func_D_ev}, \eqref{eq:func_D_od}, and \eqref{eq:func_V} in Methods, and hence influence on the key rate is expected to be small.  We thus believe that the fundamental obstacles associated with the analogue nature of the CV protocol have been settled by our approach.

To improve the presented key rate, increasing the number of states from two seems to be a promising route. Our fidelity test can be straightforwardly generalized to monitoring of such a larger constellation of signals, and we will be able to confine the adversary's attacks more tightly than in the present binary protocol. As for the proof techniques to determine the amount of privacy amplification, there are two possible directions. One is to generalize the present DV-QKD inspired approach of estimating the number of phase errors in qubits to the case of qudits. The other direction is to seek a way to combine the existing analyses \cite{four_state_Leverrier,four_state_Lutken,Devetak_Winter} of discrete modulation CV-QKD protocols, which have been reported to yield high key rates in the asymptotic regime, to our fidelity test.

In summary, we proved the security of a binary-modulated CV QKD protocol in the finite-size regime while completely circumventing the problems arising from the analogue nature of CV-QKD. We believe that it is a significant milestone toward real-world implementation of CV-QKD, which has its own advantages.

\bigskip

\noindent{\bf \large Methods}

\noindent{\bf Proof of Theorem 1 and Eq.~\eqref{eq:distance_coherent}.}
From Eq.~\eqref{eq:hetero_density}, the expectation value of $\Lambda_{m,r}(|\hat{\omega}|^2)$ when given a measured state $\rho$ is given by
\begin{align}
  &\mathbb{E} _\rho[\Lambda_{m,r}\(|\hat{\omega}|^2\)] \nonumber \\
  &=\int_{\omega\in\mathbb{C}}  \Lambda_{m,r}\(|\omega|^2\) q_{\rho}(\omega)\, d^2\omega \nonumber \\
  &= \int_0^{\infty} d\mu\, \Lambda_{m,r}(\mu) \(\int_{0}^{2\pi} \frac{d\theta}{2\pi} \bra{\sqrt{\mu}e^{i\theta}}\rho\ket{\sqrt{\mu}e^{i\theta}}\) \nonumber \\
  &= \int_0^{\infty} d\mu\, \Lambda_{m,r}(\mu) \(\sum_{n=0}^{\infty} \frac{\mu^{n}e^{-\mu}}{n!} \bra{n}\rho\ket{n}\) \nonumber \\
  & =\sum_{n=0}^{\infty}\frac{\bra{n}\rho\ket{n}I_{n,m}}{(1+r)^n},
  \label{eq:estimation_value}
\end{align}
where 
\begin{equation}
  \label{eq:I_n,m}
  I_{n,m}\coloneqq \frac{1}{n!}\int_0^\infty d\mu\ e^{-\mu}\mu^n L_m^{(1)}(\mu)
\end{equation}
for integers $n,m \geq 0$.

One can show the following three properties with regard to $I_{n,m}$:

\begin{itemize}
  \item[(i)]$I_{n,m}=0$ for $m\geq n \geq 1$. 
\end{itemize}
This results from orthogonality relations of the associated Laguerre polynomials, that is,
\begin{equation}
  \int_0^\infty L_n^{(1)}(\mu)L_m^{(1)}(\mu) \mu e^{-\mu} \, d\mu = (n+1)\delta_{n,m}.
\end{equation}
Since the polynomial $\mu^{n-1}$ can be written as a linear combination of lower order polynomials $\{L_l^{(1)}(\mu)\}_{0\leq l \leq n-1}$, $I_{n,m}$ vanishes whenever $m\geq n \geq 1$.

\begin{itemize}
  \item[(ii)] $(-1)^mI_{n,m}>0$ for $n> m \geq 0$.
\end{itemize}
This property is shown as follows. First, the associated Laguerre polynomials satisfy the following recurrence relation for $m\geq 1$ \cite{laguerre}:
\begin{equation}
  mL_m^{(1)}(\mu)=\mu\frac{dL_m^{(1)}}{d\mu}(\mu)+(m+1)L_{m-1}^{(1)}(\mu).
\end{equation}
Substituting this to Eq.~\eqref{eq:I_n,m} and using integration by parts, we have
\begin{equation}
  \label{eq:I_nm_recurrence_relation}
  I_{n,m}=\frac{n+m}{n}I_{n-1,m}-\frac{m+1}{n}I_{n-1,m-1}.
\end{equation}
for $n\geq1 $ and $m\geq 1$.
The property (ii) is then proved by induction over $m$. For $m=0$, it is true since $I_{n,0}=1>0$.
When $(-1)^{m-1} I_{n,m-1}>0$ for $n>m-1$, we can prove $(-1)^{m} I_{n,m}>0$ for $n>m$ by using Eq.~\eqref{eq:I_nm_recurrence_relation} recursively with $I_{m,m}=0$
from property (i).

\begin{itemize}
  \item[(iii)]
  $I_{0,m}=1$ for $m\geq 0$.
\end{itemize}
This also follows from property (i) and Eq.~\eqref{eq:I_nm_recurrence_relation} for $n=1$ and $m\geq 1$, 
which leads to $I_{0,m}=I_{0,0}=1$.

Properties (i), (ii), and (iii) prove Theorem 1.  Eq.~\eqref{eq:expectation_test_function} immediately follows. Because of the property (ii), $\mathbb{E}_{\rho}[\Lambda_{m,r}(\mu)]$ is always less than $c_0=F(\ket{0}\!\bra{0},\rho)$ when $m$ is odd, that is, for an arbitrary state $\rho$ and an odd integer $m$, we have
\begin{equation}
    \mathbb{E}_{\rho}[\Lambda_{m,r}(\mu)]\leq {\rm Tr}\(\rho\ket{0}\!\bra{0}\).
\end{equation}

The generalization to the fidelity to a coherent state $\ket{\beta}$ is justified in the following way.
Let $D_\beta$ be the displacement operator satisfying
\begin{equation}
    D_\beta \ket{0}\!\bra{0} D_\beta^\dagger =\ket{\beta}\!\bra{\beta},
\end{equation}
and $D^{\dagger}_{\beta}=D_{-\beta}$.  With $\tilde{\rho}\coloneqq D_\beta\rho D_\beta^\dagger$, we have $q_{\tilde{\rho}}({\omega})=q_{\rho}({\omega}-\beta)$, which implies that
\begin{align}
      \mathbb{E}_{\tilde{\rho}}[\Lambda_{m,r}(|\hat{\omega}-\beta|^2)] &= \mathbb{E}_{\rho}[\Lambda_{m,r}(|\hat{\omega}|^2)] \nonumber \\
      & \leq \bra{0}\rho\ket{0} \nonumber\\
      &= \bra{\beta}\tilde{\rho}\ket{\beta}.
    \label{eq:estimation_value_beta}
\end{align}
Replacing $\tilde{\rho}$ with $\rho$,  we obtain Eq.~\eqref{eq:distance_coherent}.

\medskip

\noindent{\bf Definition of security in the finite-size regime.}
We evaluate the secrecy of the final key as follows.
When the final key length is $\hat{N}^{\rm fin} \ge 1$, we represent Alice's final key and an adversary's quantum system as a joint state
\begin{equation}
\rho^\mathrm{fin}_{\mathrm{AE}|\hat{N}^{\rm fin}}=\sum_{z=0}^{2^{\hat{N}^{\rm fin}}-1} \mathrm{Pr}(z) \ket{z}\bra{z}_\mathrm{A} \otimes \rho^\mathrm{fin}_{\mathrm{E}|\hat{N}^{\rm fin}}(z),
\end{equation}
and define the corresponding ideal state as 
\begin{equation}
\rho^\mathrm{ideal}_{\mathrm{AE}|\hat{N}^{\rm fin}}=\sum_{z=0}^{2^{\hat{N}^{\rm fin}}-1} 2^{-\hat{N}^{\rm fin}} \ket{z}\bra{z}_\mathrm{A} \otimes 
\mathrm{Tr_A} (\rho^\mathrm{fin}_{\mathrm{AE}|\hat{N}^{\rm fin}}).
\end{equation}
Let $\|\sigma\|_1=\mathrm{Tr}\sqrt{\sigma^\dagger \sigma}$ be the trace norm of an operator $\sigma$.
We say a protocol is $\epsilon_{\rm sct}$-secret when
\begin{equation}
\frac{1}{2}\sum_{\hat{N}^{\rm fin}\geq 1} \mathrm{Pr}(\hat{N}^{\rm fin}) \|\rho^\mathrm{fin}_{\mathrm{AE}|\hat{N}^{\rm fin}} -  \rho^\mathrm{ideal}_{\mathrm{AE}|\hat{N}^{\rm fin}} \|_1
\le \epsilon_{\rm sct}
\end{equation}
holds regardless of the adversary's attack. It is known \cite{Hayashi_Tsurumaru} that if the number of phase errors is bounded as shown in 
Eq.~\eqref{eq:probability_condition}, the protocol with Eq.~\eqref{eq:final_key_length} is $\epsilon_{\rm sct}$-secret with 
$\epsilon_\mathrm{sct}=\sqrt{2}\sqrt{\epsilon+2^{-s}}$.

For correctness, we say a protocol is $\epsilon_{\rm cor}$-correct if the probability for Alice's and Bob's final key to differ is bounded by 
$\epsilon_{\rm cor}$. Our protocol achieves $\epsilon_{\rm cor}=2^{-s'}$ via the verification in Step~4.

When the above two conditions are met, the protocol becomes $\epsilon_{\rm sec}$-secure with 
$\epsilon_{\rm sec}=\epsilon_{\rm sct}+\epsilon_{\rm cor}$ in the sense of universal composability \cite{composability}.

\medskip
\noindent{\bf Derivation of the operator inequality.}
Here we construct $B(\kappa,\gamma)$ which fulfills the operator inequality \eqref{eq:operator_inequality}. 
Let us denote the supremum of the spectrum of a bounded self-adjoint operator $O$ by $\sigma_{\rm sup}(O)$.
Although $\sigma_{\rm sup}(M[\kappa,\gamma])$ would give a tightest bound, it is hard to compute numerically since system $B$ has an infinite-dimensional Hilbert space. 
Instead, we derive a looser but simpler bound.

To make use of the symmetry in the problem, we aim at bounding another operator $M'[\kappa,\gamma^{+},\gamma^{-}]$ defined by
\begin{equation}
  M'[\kappa,\gamma^{+},\gamma^{-}] \coloneqq M^{\rm suc}_{\rm ph} + \kappa \Pi_{\rm fid} - \gamma^{+}\Pi_{+} - \gamma^{-}\Pi_{-},
\end{equation}
We see that
\begin{equation}
  M[\kappa,\gamma] = M'[\kappa,0,\gamma]. \label{eq:relation_M_M'}
\end{equation}
Let us introduce projection operators by
\begin{equation}
    \Pi_{\pm,{\rm ev(od)}}\coloneqq \ket{\pm}\!\bra{\pm}_A\otimes \Pi_{{\rm ev(od)}},
\end{equation}
where $\Pi_{{\rm ev(od)}}$ is defined in the main text.
Let us further introduce orthogonal states as
\begin{align}
  \label{eq:orthogonal_four_states_2}
  \ket{\phi_{\rm err}}_{AB}&\coloneqq \ket{+}_{A}\otimes \Pi_{\rm od}\ket{\beta}_B + \ket{-}_{A}\otimes \Pi_{\rm ev}\ket{\beta}_B, \\
  \label{eq:orthogonal_four_states}
  \ket{\phi_{\rm cor}}_{AB}&\coloneqq \ket{-}_{A}\otimes\Pi_{\rm od}\ket{\beta}_B + \ket{+}_{A}\otimes\Pi_{\rm ev}\ket{\beta}_B.
\end{align}
One can check that
\begin{equation}
  \Pi_{\rm fid}= \ket{\phi_{\rm err}}\!\bra{\phi_{\rm err}} + \ket{\phi_{\rm cor}}\!\bra{\phi_{\rm cor}}
\end{equation}
holds by using $(\Pi_{\rm ev}-\Pi_{\rm od})\ket{\beta}_B=\ket{-\beta}_B$.
We can then decompose $M'[\kappa,\gamma^{+},\gamma^{-}]$ into a direct sum of two operators as follows:
\begin{equation}
    \label{eq:direct_sum}
    M'[\kappa,\gamma^{+},\gamma^{-}] =  M_{\rm err} \oplus M_{\rm cor},
\end{equation} 
where 
\begin{align}
    M_{\rm err} &\coloneqq M_{\rm ph}^{\rm suc}+\kappa\ket{\phi_{\rm err}}\!\bra{\phi_{\rm err}}-\gamma^{+}\Pi_{+,{\rm od}}-\gamma^{-}\Pi_{-,{\rm ev}}, \label{eq:M_err_definition} \\
    M_{\rm cor} &\coloneqq \kappa\ket{\phi_{\rm cor}}\!\bra{\phi_{\rm cor}}-\gamma^{+}\Pi_{+,{\rm ev}}-\gamma^{-}\Pi_{-,{\rm od}}. \label{eq:M_cor_definition}
\end{align}

We define an orthonormal basis $\{\ket{e_{\rm ev}^{(j)}}, \ket{e_{\rm od}^{(j)}}\}_{j=1}^{\infty}$ through the following equations:
\begin{align}
    & \sqrt{C_{\rm ev(od)}}\ket{e_{\rm ev(od)}^{(1)}}_B = \Pi_{\rm ev(od)}\ket{\beta}_B,\\
    \begin{split}
        \label{eq:B_basis_2}
        &M_{\rm ev(od)}^{\rm suc}\ket{e_{\rm ev(od)}^{(1)}}_B \\
        & \ = D_{\rm ev(od)}\ket{e_{\rm ev(od)}^{(1)}}_B+\sqrt{V_{\rm ev(od)}}\ket{e_{\rm ev(od)}^{(2)}}_B,
    \end{split}\\
    \label{eq:B_basis_sum}
    &\sum_{j=1}^{\infty}\ket{e_{\rm ev(od)}^{(j)}}\!\bra{e_{\rm ev(od)}^{(j)}}=\Pi_{{\rm ev(od)}}.
\end{align}
The normalization factors in Eq.~\eqref{eq:B_basis_2} are explicitly given by 
\begin{align}
    C_{\rm ev} &\coloneqq \bra{\beta}\Pi_{\rm ev}\ket{\beta} = e^{-|\beta|^2}\cosh|\beta|^2, \label{eq:definition_C}\\
    C_{\rm od} &\coloneqq \bra{\beta}\Pi_{\rm od}\ket{\beta} = e^{-|\beta|^2}\sinh|\beta|^2, \\
    \label{eq:definition_D} 
    D_{\rm ev(od)} &\coloneqq C_{\rm ev(od)}^{-1}\bra{\beta} M_{{\rm ev(od)}}^{\rm suc}\ket{\beta},\\
    \label{eq:definition_V}
    V_{\rm ev(od)} &\coloneqq C_{\rm ev(od)}^{-1}\bra{\beta}\bigl(M_{{\rm ev(od)}}^{\rm suc}\bigr)^2\ket{\beta} - D_{\rm ev(od)}^2.
\end{align}

{\noindent These quantities can be numerically computed by integration through Eq.~\eqref{eq:parity_measurement}.}
We further define the following projectors:
\begin{align}
    \label{eq:definition_projector_12}
    \Pi_{\rm ev(od)}^{(j)}&\coloneqq \ket{e_{\rm ev(od)}^{(j)}}\!\bra{e_{\rm ev(od)}^{(j)}} \quad (j=1,2),\\
    \label{eq:definition_projector_3}
    \Pi_{\rm ev(od)}^{(\geq 3)}&\coloneqq \Pi_{\rm ev(od)} - \Pi_{\rm ev(od)}^{(1)} -\Pi_{\rm ev(od)}^{(2)}.
\end{align}

Since Eq.~\eqref{eq:definition_D} implies  $\Pi_{\rm od}^{(1)}M_{\rm od}^{\rm suc}\Pi_{\rm od}^{(\geq3)}=0$, we have
\begin{equation}
    \label{eq:operator_transform_1}
    \begin{split}
        M_{\rm od}^{\rm suc} &= \Pi_{\rm od}^{(1)}M_{\rm od}^{\rm suc}\Pi_{\rm od}^{(1)} + \Pi_{\rm od}^{(\geq 2)}M_{\rm od}^{\rm suc}\Pi_{\rm od}^{(\geq 2)}\\
        & \qquad + \Pi_{\rm od}^{(1)}M_{\rm od}^{\rm suc}\Pi_{\rm od}^{(2)}+\Pi_{\rm od}^{(2)}M_{\rm od}^{\rm suc}\Pi_{\rm od}^{(1)}.
    \end{split}
\end{equation}
The last term is bounded as
\begin{equation}
    \Pi_{\rm od}^{(\geq 2)}M_{\rm od}^{\rm suc}\Pi_{\rm od}^{(\geq 2)}\leq \Pi_{\rm od}^{(\geq 2)}
    \label{eq:bound_M_od}
\end{equation}
since $M_{\rm od}^{\rm suc} \leq \bm{1}_{B}$.
Combining Eqs.~\eqref{eq:definition_D}, \eqref{eq:operator_transform_1}, and \eqref{eq:bound_M_od}, we have
\begin{equation}
    \begin{split}
        \label{eq:upperbound_operator_cor_1}
        &M_{\rm od}^{\rm suc}-\gamma^{+}\Pi_{\rm od} \\
        & \leq (D_{\rm od}-\gamma^{+})\ket{e_{\rm od}^{(1)}}\!\bra{e_{\rm od}^{(1)}} +(1-\gamma^{+})\ket{e_{\rm od}^{(2)}}\!\bra{e_{\rm od}^{(2)}} \\
        &\quad + \sqrt{V_{\rm od}}(\ket{e_{\rm od}^{(2)}}\!\bra{e_{\rm od}^{(1)}}+\ket{e_{\rm od}^{(1)}}\!\bra{e_{\rm od}^{(2)}}) + (1-\gamma^{+})\Pi_{\rm od}^{(\geq 3)}.
    \end{split}
\end{equation}
In the same way, by replacing $+\leftrightarrow -$ and ${\rm od}\leftrightarrow {\rm ev}$, we have
\begin{equation}
    \begin{split}
        \label{eq:upperbound_operator_cor_2}
        &M_{\rm ev}^{\rm suc}-\gamma^{-}\Pi_{\rm ev}\\
        & \leq (D_{\rm ev}-\gamma^{-})\ket{e_{\rm ev}^{(1)}}\!\bra{e_{\rm ev}^{(1)}} +(1-\gamma^{-})\ket{e_{\rm ev}^{(2)}}\!\bra{e_{\rm ev}^{(2)}}\\
        &\ \quad +\sqrt{V_{\rm ev}}(\ket{e_{\rm ev}^{(2)}}\!\bra{e_{\rm ev}^{(1)}}+\ket{e_{\rm ev}^{(1)}}\!\bra{e_{\rm ev}^{(2)}}) + (1-\gamma^{-})\Pi_{\rm ev}^{(\geq 3)}.
    \end{split}
\end{equation}
Using Eqs.~\eqref{eq:upperbound_operator_cor_1} and \eqref{eq:upperbound_operator_cor_2}, we can bound the operator $M_{\rm err}$ in Eq.~\eqref{eq:M_err_definition} as
\begin{equation}
        \label{eq:upperbound_M_err}
        M_{\rm err}\leq M_{\rm err}^{\text{r-4}}[\kappa,\gamma^{+},\gamma^{-}] \oplus (1-\gamma^{+})\Pi_{\rm +, od}^{(\geq 3)} \oplus (1-\gamma^{-})\Pi_{\rm -, ev}^{(\geq 3)},
\end{equation}
with a rank-4 operator 
\begin{equation}
    \begin{split}
        \label{eq:definition_M_err^fin}
        &M_{\rm err}^{\text{r-4}}[\kappa,\gamma^+,\gamma^-] \\
        &\coloneqq \kappa\ket{\phi_{\rm err}}\!\bra{\phi_{\rm err}}\\
        &\quad +(D_{\rm od}-\gamma^{+})\ket{e_{+,{\rm od}}^{(1)}}\!\bra{e_{+,{\rm od}}^{(1)}} + (1-\gamma^{+})\ket{e_{+,{\rm od}}^{(2)}}\!\bra{e_{+,{\rm od}}^{(2)}}\\
        &\ \quad +\sqrt{V_{\rm od}}(\ket{e_{+,{\rm od}}^{(2)}}\!\bra{e_{+,{\rm od}}^{(1)}}+\ket{e_{+,{\rm od}}^{(1)}}\!\bra{e_{+,{\rm od}}^{(2)}})\\
        &\ \ \quad +(D_{\rm ev}-\gamma^{-})\ket{e_{-,{\rm ev}}^{(1)}}\!\bra{e_{-,{\rm ev}}^{(1)}} +(1-\gamma^{-})\ket{e_{-,{\rm ev}}^{(2)}}\!\bra{e_{-,{\rm ev}}^{(2)}}\\
        &\ \ \ \quad +\sqrt{V_{\rm ev}}(\ket{e_{-,{\rm ev}}^{(2)}}\!\bra{e_{-,{\rm ev}}^{(1)}}+\ket{e_{-,{\rm ev}}^{(1)}}\!\bra{e_{-,{\rm ev}}^{(2)}}),
    \end{split}
\end{equation}
where 
\begin{equation}
  \Pi_{\pm, {\rm od(ev)}}^{(\geq 3)}\coloneqq \ket{\pm}\!\bra{\pm}_A\otimes\Pi_{{\rm od(ev)}}^{(\geq 3)}
\end{equation}
and
\begin{equation}
    \ket{e_{\rm \pm,od(ev)}^{(j)}}_{AB}\coloneqq \ket{\pm}_A\ket{e_{\rm od(ev)}^{(j)}}_B.
\end{equation}
Using the basis $\{\ket{e_{+,{\rm od}}^{(2)}},\ket{e_{+,{\rm od}}^{(1)}},\ket{e_{-,{\rm ev}}^{(1)}},\ket{e_{-,{\rm ev}}^{(2)}}\}$, we have a matrix representation of $M_{\rm err}^{\text{r-4}}[\kappa,\gamma^+,\gamma^-]$ defined in Eq.~\eqref{eq:definition_M_err^fin} as follows:
\begin{equation}
    \begin{bmatrix}
        1 - \gamma^{+} & \sqrt{V_{\rm od}} &  & \\
        \sqrt{V_{\rm od}} & \kappa\, C_{\rm od} + D_{\rm od}\!-\gamma^{+}\! & \kappa\sqrt{C_{\rm od}\, C_{\rm ev}} & \\
         & \kappa\sqrt{C_{\rm od}\, C_{\rm ev}}, & \kappa\, C_{\rm ev} + D_{\rm ev}\!-\gamma^{-}\!& \sqrt{V_{\rm ev}}\\
         & & \sqrt{V_{\rm ev}} & 1 -\gamma^{-}
    \end{bmatrix}.
\end{equation}

As for $M_{\rm cor}$, it has a simpler decomposition:
\begin{equation}
        \label{eq:upperbound_M_cor}
        M_{\rm cor}= M_{\rm cor}^{\text{r-2}}[\kappa,\gamma^+,\gamma^-] \oplus (-\gamma^{+})\Pi_{\rm +, ev}^{(\geq 2)} \oplus (-\gamma^{-})\Pi_{\rm -, od}^{(\geq 2)},
\end{equation}
where a rank-2 operator $M_{\rm cor}^{\text{r-2}}[\kappa,\gamma^+,\gamma^-]$ is given by,
\begin{align}
  &M_{\rm cor}^{\text{r-2}}[\kappa,\gamma^+,\gamma^-] \nonumber \\
   &\coloneqq  \kappa\ket{\phi_{\rm cor}}\!\bra{\phi_{\rm cor}} -\gamma^{+}\ket{e_{+,{\rm ev}}^{(1)}}\!\bra{e_{+,{\rm ev}}^{(1)}} -\gamma^{-}\ket{e_{-,{\rm od}}^{(1)}}\!\bra{e_{-,{\rm od}}^{(1)}} \nonumber \\
    &= \begin{bmatrix} \kappa\, C_{\rm ev} - \gamma^+ & \kappa \sqrt{C_{\rm ev}\, C_{\rm od}}\\
        \kappa \sqrt{C_{\rm ev}\, C_{\rm od}} & \kappa\, C_{\rm od} - \gamma^- \end{bmatrix},
  \label{eq:M_cor_fin}
\end{align}

{\noindent where we used the basis $\{\ket{e_{+,{\rm ev}}^{(1)}},\ket{e_{-,{\rm od}}^{(1)}}\}$ for the matrix representation.}

Finally, from Eqs.~\eqref{eq:relation_M_M'}, \eqref{eq:direct_sum}, \eqref{eq:upperbound_M_err}, and \eqref{eq:upperbound_M_cor}, we obtain an upper bound on $\sigma_{\rm sup}(M[\kappa,\gamma])$ as
\begin{equation}
        \label{eq:upperbound_supremum}
        \sigma_{\rm sup}(M[\kappa,\gamma])
        =\sigma_{\rm sup}(M'[\kappa,0,\gamma]) \leq B(\kappa,\gamma),
\end{equation}
with 
\begin{equation}
  \begin{split}
  &B(\kappa,\gamma) \\
  &\coloneqq {\rm max}\left\{\sigma_{\rm sup}\bigl(M_{\rm err}^{\text{r-4}}[\kappa,0,\gamma]\bigr),\sigma_{\rm sup}\bigl(M_{\rm cor}^{\text{r-2}}[\kappa,0,\gamma]\bigr), 1 \right\},
  \end{split}
  \label{eq:def_of_B}
\end{equation}
where we used $\gamma \geq 0$.
Since $M_{\rm cor}^{\text{r-4}}$ and $M_{\rm err}^{\text{r-2}}$ are four-dimensional and two-dimensional matrices, their largest eigenvalues can be numerically calculated.

\medskip

{\noindent \bf Derivation of the finite size bound.}
Here we construct the function $U(\hat{F},\hat{N}^{\rm trash})$ to satisfy Eq.~\eqref{eq:probability_condition} in the modified scenario.  For that, we will first derive Eq.~\eqref{eq:bound_for_T}.
In the modified scenario, we define the following random variables labeled by the number $i$ of the round;
\begin{itemize}
  \item[(i)] $\hat{N}_{\rm ph}^{{\rm suc}, (i)}$ is defined to be unity only when ``signal'' is chosen in the $i$-th round, the detection is a ``success'', and a pair of outcomes $(a', b')$ is $(+,{\rm odd})$ or $(-,{\rm even})$.  Otherwise, $\hat{N}_{\rm ph}^{{\rm suc}, (i)}=0$. We have
  \begin{equation}
    \qquad \ \ \hat{N}_{\rm ph}^{{\rm suc}, (i)} =\begin{cases}
    1& \bigl({\rm signal,\ success,} \\ &\qquad  (+,{\rm odd}) \text{ or } (-, {\rm even})\bigr)\\
      0& ({\rm otherwise})
    \end{cases},
  \end{equation}
  and $\hat{N}_{\rm ph}^{\rm suc} = \sum_{i=1}^{N}\hat{N}_{\rm ph}^{{\rm suc}, (i)}$. 

  \item[(ii)]
  $\hat{F}^{(i)}$ is defined to be $\Lambda_{m,r}(|\hat{\omega} -(-1)^a\beta|^2)$ when ``test'' is chosen in the $i$-th round. We have 
  \begin{equation}
    \qquad \ \hat{F}^{(i)} =\begin{cases}
      \Lambda_{m,r}(|\hat{\omega} -(-1)^a\beta|^2)& ({\rm test})\\
      0& ({\rm otherwise})
    \end{cases},
  \end{equation}
  and $\hat{F} = \sum_{i=1}^{N}\hat{F}^{(i)}$.

  \item[(iii)] 
  $\hat{Q}_{-}^{(i)}$ is defined to be unity only when ``trash'' is chosen in the $i$-th round and $a'=-$.  Otherwise, $\hat{Q}_{-}^{(i)}=0$.  We have
  \begin{equation}
    \hat{Q}_{-}^{(i)} =\begin{cases}
      1& ({\rm trash},\ -)\\
      0& ({\rm otherwise})
    \end{cases},
  \end{equation}
  and $\hat{Q}_{-}=\sum_{i=1}^{N}\hat{Q}_{-}^{(i)}$.

  \item[(iv)]
  We also define 
  \begin{equation}
    \qquad \ \ \hat{T}^{(i)}\coloneqq p_{\rm sig}^{-1}\hat{N}_{\rm ph}^{{\rm suc},(i)}+p_{\rm test}^{-1}\kappa\hat{F}^{(i)}-p_{\rm trash}^{-1}\gamma\hat{Q}_{-}^{(i)},
  \end{equation}
  which leads to $\hat{T}[\kappa,\gamma]=\sum_{i=1}^{N}\hat{T}^{(i)}$.
\end{itemize}

We will make use of Azuma's inequality \cite{Azuma_ineq}. We define stochastic processes $\{\hat{X}^{(k)}\}_{k=0,\ldots, N}$ and $\{\hat{Y}^{(k)}\}_{k=1,\ldots, N}$ as follows:
\begin{align}
    \label{eq:X_0}
    \hat{X}^{(0)} &\coloneqq 0,\\
    \label{eq:X_1}
    \hat{X}^{(k)} &\coloneqq \sum_{i=1}^{k} \bigl(\hat{T}^{(i)} - \hat{Y}^{(i)}\bigr) \quad (k\geq 1),\\
    \label{eq:A_1}
    \hat{Y}^{(k)} &\coloneqq \mathbb{E}\bigl[\hat{T}^{(k)}\bigr|\hat{X}^{<k}\bigr],
\end{align}

{\noindent where $\hat{X}^{<k}\coloneqq (\hat{X}^{(0)},\hat{X}^{(1)},\ldots, \hat{X}^{(k-1)})$.}
Note that $\hat{Y}^{(k)}$ is a constant when conditioned on $\hat{X}^{<k}$.  Such a sequence $\{\hat{Y}^{(k)}\}_{k=1,2,\ldots}$ is called a predictable process with regards to $\{\hat{X}^{(k)}\}$.
Since $\hat{T}^{(i)}$ is bounded for any $i$ and $\{\hat{X}^{(k)}\}_{k=0,1,\ldots}$ is a martingale, we can apply Azuma's inequality.

\medskip

\noindent{\it Proposition {\rm (Generalized Azuma's inequality \cite{refined_Azuma_ineq1, refined_Azuma_ineq2}):}} Suppose $\{\hat{X}^{(k)}\}_{k=0,1,\ldots}$ is a martingale which satisfies
\begin{equation}
    \label{eq:Azuma_application_condition}
  -\hat{Y}^{(k)} + c_{\min} \leq \hat{X}^{(k)}-\hat{X}^{(k-1)} \leq -\hat{Y}^{(k)} + c_{\max},
\end{equation}
for constants $c_{\min}$ and $c_{\max}$, and a predictable process $\{\hat{Y}^{(k)}\}_{k=1,2,\ldots}$ with regards to $\{\hat{X}^{(k)}\}$, i.e., $\hat{Y}^{(k)}$ is constant when conditioned on $\hat{X}^{<k}$.
Then, for all positive integers $N$ and all positive reals $\delta$,
\begin{equation}
  {\rm Pr}[\hat{X}^{(N)}-\hat{X}^{(0)}\geq\delta]\leq {\rm exp}\left(-\frac{2\delta^2}{(c_{\max}-c_{\min})^2 N}\right).
\end{equation}
\medskip

{\noindent We define constants $c_{\min}$ and $c_{\max}$ as follows.}
In each round, at most one of $\hat{N}_{\rm ph}^{{\rm suc}, (i)}$, $\hat{F}^{(i)}$, and $\hat{Q}_{-}^{(i)}$ takes non-zero value; $\hat{N}_{\rm ph}^{{\rm suc}, (i)}$ and $\hat{Q}_{-}^{(i)}$ are either zero or unity, and ${\rm min}\Lambda_{m,r}\leq \hat{F}^{(i)} \leq {\rm max}\Lambda_{m,r}$.  Since $\kappa,\gamma \geq 0$, Eq.~\eqref{eq:Azuma_application_condition} holds when $c_{\min}$ and $c_{\max}$ are defined as 
\begin{align}
  c_{\min} &\coloneqq {\rm min}\Bigl(p_{\rm test}^{-1}\kappa\; {\rm min}\Lambda_{m,r},\ -p_{\rm trash}^{-1}\gamma,\ 0\Bigr), \\
  c_{\max} &\coloneqq {\rm max}\Bigl(p_{\rm sig}^{-1},\ p_{\rm test}^{-1}\kappa\; {\rm max}\Lambda_{m,r},\ 0\Bigr).
\end{align}
With $c_{\min}$ and $c_{\max}$ defined as above, we further define 
\begin{equation}
  \delta_1(\epsilon) \coloneqq (c_{\max}-c_{\min})\sqrt{\frac{N}{2}{\rm ln}\left(\frac{1}{\epsilon}\right)}.
\end{equation}
Setting $\delta=\delta_1(\epsilon/2)$ in the proposition, we conclude that 
\begin{equation}
        \hat{T}[\kappa,\gamma]\leq \sum_{i=1}^{N}\hat{Y}^{(i)} + \delta_1(\epsilon/2)
        \label{eq:result_azuma}
\end{equation}
holds with a probability no smaller than $1-\epsilon/2$.

Next, we will construct a deterministic bound on $\hat{Y}^{(i)}$. 
Let $\rho^{(i)}_{AB}$ be the state of Alice's $i$-th qubit and Bob's $i$-th pulse conditioned on $\hat{X}^{<i}$.  
Then, using the same argument as that has lead to Eqs.~\eqref{eq:expectation_n_ph}--\eqref{eq:expectation_f}, we have
\begin{align}
    \label{eq:trace_signal}
    \mathbb{E}\Bigl[\hat{N}_{\rm ph}^{\mathrm{suc}, (i)}\Bigr|\hat{X}^{<i}\Bigr]&= p_{\rm sig} {\rm Tr}\(\rho^{(i)}_{AB} M_{\rm ph}^{\rm suc}\),\\
    \label{eq:trace_trash}
    \mathbb{E}\Bigl[\hat{Q}_{-}^{(i)}\Bigr|\hat{X}^{<i}\Bigr]&= p_{\rm trash}{\rm Tr}\(\rho^{(i)}_{AB} \Pi_{-}\), \\
    \label{eq:trace_test}
    \mathbb{E}[\hat{F}^{(i)}|\hat{X}^{<i}]&\leq p_{\rm fid} {\rm Tr}\(\rho^{(i)}_{AB} \Pi_{\rm fid}\),
\end{align}
and thus
\begin{equation}
  \hat{Y}^{(i)} \leq {\rm Tr}\(\rho^{(i)}_{AB}M[\kappa,\gamma]\),
\end{equation}
where $M[\kappa,\gamma]$ is defined in Eq.~\eqref{eq:def_of_M}.
Using the operator inequality \eqref{eq:operator_inequality}, we obtain a bound independent of $i$ as
\begin{equation}
    \label{eq:Trace_M_upper_bound}
    \hat{Y}^{(i)} \leq B(\kappa,\gamma).
\end{equation}
Combining this with Eq.~\eqref{eq:result_azuma} proves Eq.~\eqref{eq:bound_for_T}.

The function $\delta_2(\epsilon/2;\hat{N}^{\rm trash})$ satisfying the bound \eqref{eq:bound_Q} on $\hat{Q}_{-}$ can be derived from the fact that $\mathrm{Pr}[\hat{Q}_{-}| \hat{N}^{\mathrm{trash}}]$ is a binomial distribution.
The following inequality thus holds for any positive integer $n$ and a real $\delta$ with $0<\delta<(1-q_{-})n$ (Chernoff bound):
\begin{equation}
  \begin{split}
  &{\rm Pr}\Bigl[\hat{Q}_{-} - q_{-} n \geq \delta\Bigr|\hat{N}^{\rm trash}=n\Bigr]\\ 
  & \qquad \quad \leq \exp\left[- n D\bigl(q_{-} + \delta/n\bigr\|q_{-}\bigr)\right],
  \end{split}
  \label{eq:Q_minus_domain}
\end{equation}
where 
\begin{equation}
  D(x\|y)\coloneqq x{\log}\frac{x}{y}+(1-x){\log}\frac{1-x}{1-y}
\end{equation}
is the Kullback-Leibler divergence.
On the other hand, for any non-negative integer $n$, we always have
\begin{equation}
  \mathrm{Pr}\Bigl[\hat{Q}_{-} - q_{-} n \leq (1-q_{-})n\Bigr|\hat{N}^{\rm trash}=n\Bigr] = 1. \label{eq:bound_q_minus_trivial}
\end{equation}

{\noindent Therefore, for any non-negative integer $n$, by defining $\delta_2(\epsilon;n)$ which satisfies}
\begin{equation}
  \begin{cases}
  \exp\left[-n D\bigl(q_{-} +\delta_2(\epsilon;n)/n \bigr\| q_{-} \bigr)\right] = \epsilon & (\epsilon > q_{-}^n) \\
  \delta_2(\epsilon;n) = (1-q_{-})n & (\epsilon \leq q_{-}^n)
  \end{cases},
\end{equation}
and by combining Eq.~\eqref{eq:Q_minus_domain} and \eqref{eq:bound_q_minus_trivial}, we conclude that Eq.~\eqref{eq:bound_Q} holds with a probability no smaller than $1-\epsilon/2$.

Combining Eq.~\eqref{eq:bound_for_T} and Eq.~\eqref{eq:bound_Q}, we obtain Eq.~\eqref{eq:probability_condition} by setting
\begin{equation}
    \begin{split}
      &U(\hat{F},\hat{N}^{\rm trash}) \\
      &\coloneqq p_{\rm sig}N B(\kappa, \gamma)+ p_{\rm sig}\delta_1(\epsilon/2) \\
      &\qquad -\frac{p_{\rm sig}}{p_{\rm test}} \kappa \hat{F} + \frac{p_{\rm sig}}{p_{\rm trash}} \gamma \(q_{-}\hat{N}^{\rm trash}+\delta_2(\epsilon/2;\hat{N}^{\rm trash})\)
    \end{split}
\end{equation}
which holds with a probability no smaller than $1-\epsilon$ (Union bound).

\medskip
\noindent{\bf Model of the quantum channel and measurement for the calculation of key rates.}
In what follows, we normalize quadrature $x$ such that a coherent state $\ket{\omega}$ has expectation $\braket{x}= \mathrm{Re}(\omega)$ and variance  $\braket{(\Delta x)^2}=1/4$.
The wave function for $\omega=\omega_{R} +i  \omega_{I}$ is given by
\begin{equation}
  \braket{x|\omega}=\left(\frac{2}{\pi}\right)^{\frac{1}{4}} \exp\bigl[ -(x-\omega_{R})^2+ 2i \omega_{I} x  - i  \omega_{R} \omega_{I}\bigr].
\end{equation}

For the simulation of the key rate $G$, we assume that the communication channel and Bob's detection apparatus can be modeled by a pure loss channel followed by random displacement, that is, the states which Bob receives are given by
\begin{equation}
  \label{eq:model_received_state}
  \rho_{\rm model}^{(a)} \coloneqq \int_{\mathbb{C}} p_\xi(\gamma)\ket{(-1)^a\sqrt{\eta\mu}+\gamma}\!\bra{(-1)^a\sqrt{\eta\mu}+\gamma}d^{2}\gamma,
\end{equation}
where $\eta$ is the transmissivity of the pure loss channel and $p_\xi(\gamma)$ is given by
\begin{equation}
  p_\xi(\gamma)\coloneqq \frac{2}{\pi\xi}e^{-2|\gamma|^2/\xi}.
\end{equation}
The parameter $\xi$ is the excess noise relative to the vacuum, namely,
\begin{equation}
    \Braket{(\varDelta x)^2}_{\rm \rho_{\rm model}^{(a)}} = (1+\xi)/4.
\end{equation}

We assume that Bob sets $\beta = \sqrt{\eta\mu}$ for the fidelity test.
The actual fidelity between Bob's objective state $\ket{(-1)^a\sqrt{\eta\mu}}$ and the model state $\rho_{\rm model}^{(a)}$ is given by 
\begin{align}
  \label{eq:model_actual_fidelity}
  &F(\rho_{\rm model}^{(a)},\ket{(-1)^a\sqrt{\eta\mu}}\!\bra{(-1)^a\sqrt{\eta\mu}})\nonumber \\
  &= \int_{\mathbb{C}} p_\xi(\gamma)|\braket{(-1)^a\sqrt{\eta\mu}|(-1)^a\sqrt{\eta\mu}-\gamma}|^2d\gamma \nonumber\\
  &=\frac{1}{1+\xi/2}.
\end{align}
For the acceptance probability of Bob's measurement in the signal rounds, we assume $f_{\rm suc}(x)=\Theta(|x|-x_{\rm th})$, a step function with the threshold $x_{\rm th}>0$.
In this case, the quantities defined in Eqs.~\eqref{eq:definition_D} and \eqref{eq:definition_V} are given by
\begin{align}
        D_{\rm ev} &= \int_{0}^{\infty} 2C_{\rm ev}^{-1} f_{\rm suc}(x) \, \bigl|\bra{x}\Pi_{\rm ev}\ket{\beta}\bigr|^2 dx \label{eq:func_D_ev}\\
        \begin{split}
        &= \frac{1}{4C_{\rm ev}}\Bigl[{\rm erfc}\bigl(\sqrt{2}(x_{\rm th} - \beta)\bigr) + {\rm erfc}\bigl(\sqrt{2}(x_{\rm th} + \beta)\bigr)  \\
        & \hspace{0.45\linewidth} + 2e^{-2\beta^2}{\rm erfc}\bigl(\sqrt{2}x_{\rm th}\bigr)\Bigr], 
    \end{split}\\
    D_{\rm od} &= \int_{0}^{\infty} 2C_{\rm od}^{-1} f_{\rm suc}(x) \, \bigl|\bra{x}\Pi_{\rm od}\ket{\beta}\bigr|^2 dx  \label{eq:func_D_od} \\
    \begin{split}
       &= \frac{1}{4C_{\rm od}}\Bigl[{\rm erfc}\bigl(\sqrt{2}(x_{\rm th} - \beta)\bigr) + {\rm erfc}\bigl(\sqrt{2}(x_{\rm th} + \beta)\bigr)  \\
        & \hspace{0.45\linewidth}  - 2e^{-2\beta^2}{\rm erfc}\bigl(\sqrt{2}x_{\rm th}\bigr)\Bigr], 
    \end{split}\\
  \begin{split}
    V_{\rm ev(od)} &= \int_{0}^{\infty} 2C_{\rm ev(od)}^{-1}\bigl(f_{\rm suc}(x)\bigr)^2 \bigl|\bra{x}\Pi_{\rm ev(od)}\ket{\beta}\bigr|^2 dx  \\
    & \hspace{0.6\linewidth} - D_{\rm ev(od)}^2 
  \end{split} \label{eq:func_V}\\
    &= D_{\rm ev(od)} - D_{\rm ev(od)}^2,
\end{align}
where $\beta = \sqrt{\eta\mu}$ and the complementary error function ${\rm erfc}(x)$ is defined as
\begin{equation}
    \label{eq:complementary_error_function}
    {\rm erfc}(x)\coloneqq \frac{2}{\sqrt{\pi}}\int_x^\infty dt\ e^{-t^2}.
\end{equation}
For the derivation of Eq.~\eqref{eq:func_V}, we used the fact that $\Pi_{\rm ev} + \Pi_{\rm od}=\bm{1}$ and $(\Pi_{\rm ev} - \Pi_{\rm od})\ket{\beta}=\ket{-\beta}$.

\allowdisplaybreaks[1]
We assume that the number of ``success'' signal rounds $\hat{N}^{\rm suc}$ is equal to its expectation value,
\begin{align}
    \mathbb{E}[\hat{N}^{\rm suc}]&= \(\int_{-\infty}^{\infty} f(|x|) \bra{x}\rho_{\rm model}^{(a)}\ket{x} dx \) p_{\rm sig}N \nonumber \\
    &=p_{\rm sig}N (P^+ + P^-),
\end{align}
where 
\begin{align}
    P^{\pm} &\coloneqq \int_{x_{\rm th}}^{\infty}  \bra{\pm (-1)^{a} x}\rho_{\rm model}^{(a)}\ket{\pm (-1)^{a}x} dx \nonumber\\
    &= \frac{1}{2}\,{\rm erfc}\!\((x_{\rm th} \mp \sqrt{\eta\mu}) \sqrt{\frac{2}{1+\xi}}\).
\end{align}
We also assume that the number of test rounds $\hat{N}^{\rm test}$ is equal to $p_{\rm test}N $ and the number of trash rounds $\hat{N}^{\rm trash}$ is equal to $p_{\rm trash}N$.  The test outcome $\hat{F}$ is assumed to be equal to its expectation value $\mathbb{E}[\hat{F}]$, which is given by
\begin{align}
    &\mathbb{E}[\hat{F}] \nonumber \\
    &= p_{\rm test}N\, \mathbb{E}_{\rho_{\rm model}^{(a)}}[\Lambda_{m,r}(|\hat{\omega} -(-1)^a\sqrt{\eta\mu}|^2)] \nonumber\\
    &= p_{\rm test}N \int_{\mathbb{C}}\frac{d^2\omega}{\pi}\bra{\omega}\rho_{\rm model}^{(a)}\ket{\omega}\Lambda_{m,r}(|\omega -(-1)^a\sqrt{\eta\mu}|^2) \nonumber \\
    &=\frac{p_{\rm test}N}{1+\xi/2}\left[1 - (-1)^{m+1}\left(\frac{\xi/2}{1 + r(1+\xi/2)}\right)^{m + 1}\right] . \\
    &\nonumber
\end{align} 

Under these assumptions, the key rate $\hat{G}$ for each transmissivity $\eta$ is optimized over two coefficients $(\kappa,\gamma)$ and four protocol parameters $(\mu,x_{\rm th},p_{\rm sig},p_{\rm test})$ as discussed in the main part.  The cost of bit error correction $H_{\rm EC}$ is assumed to be $1.1 \times \hat{N}^{\rm suc} h(e_{\rm bit})$,
where the bit error rate $e_{\rm bit}$ is given by
\begin{equation}
    e_{\rm bit} = \frac{P^-}{P^+ + P^-}.
\end{equation}

\noindent{\bf \large References}


\begin{thebibliography}{10}
\expandafter\ifx\csname url\endcsname\relax
  \def\url#1{\texttt{#1}}\fi
\expandafter\ifx\csname urlprefix\endcsname\relax\def\urlprefix{URL }\fi
\providecommand{\bibinfo}[2]{#2}
\providecommand{\eprint}[2][]{\url{#2}}

\bibitem{BB84}
\bibinfo{author}{Bennett, C.~H.} \& \bibinfo{author}{Brassard, G.}
\newblock \bibinfo{title}{Quantum cryptography: Public key distribution and
  coin tossing}.
\newblock In \emph{\bibinfo{booktitle}{Proceedings of IEEE International
  Conference on Computers, Systems, and Signal Processing}},
  \bibinfo{pages}{175} (\bibinfo{address}{India}, \bibinfo{year}{1984}).

\bibitem{B92}
\bibinfo{author}{Bennett, C.~H.}
\newblock \bibinfo{title}{Quantum cryptography using any two nonorthogonal
  states}.
\newblock \emph{\bibinfo{journal}{Physical Review Letters}}
  \textbf{\bibinfo{volume}{68}}, \bibinfo{pages}{3121} (\bibinfo{year}{1992}).

\bibitem{Ralph1999}
\bibinfo{author}{Ralph, T.~C.}
\newblock \bibinfo{title}{Continuous variable quantum cryptography}.
\newblock \emph{\bibinfo{journal}{Physical Review A}}
  \textbf{\bibinfo{volume}{61}}, \bibinfo{pages}{010303}
  (\bibinfo{year}{1999}).

\bibitem{Hillery2000}
\bibinfo{author}{Hillery, M.}
\newblock \bibinfo{title}{Quantum cryptography with squeezed states}.
\newblock \emph{\bibinfo{journal}{Physical Review A}}
  \textbf{\bibinfo{volume}{61}}, \bibinfo{pages}{022309}
  (\bibinfo{year}{2000}).

\bibitem{GG02}
\bibinfo{author}{Grosshans, F.} \& \bibinfo{author}{Grangier, P.}
\newblock \bibinfo{title}{Continuous variable quantum cryptography using
  coherent states}.
\newblock \emph{\bibinfo{journal}{Physical Review Letters}}
  \textbf{\bibinfo{volume}{88}}, \bibinfo{pages}{057902}
  (\bibinfo{year}{2002}).

\bibitem{Eriksson2020}
\bibinfo{author}{Eriksson, T.~A.} \emph{et~al.}
\newblock \bibinfo{title}{Wavelength division multiplexing of 194 continuous
  variable quantum key distribution channels}.
\newblock \emph{\bibinfo{journal}{Journal of Lightwave Technology}}
  \textbf{\bibinfo{volume}{38}}, \bibinfo{pages}{2214--2218}
  (\bibinfo{year}{2020}).

\bibitem{Huang2015}
\bibinfo{author}{Huang, D.} \emph{et~al.}
\newblock \bibinfo{title}{Continuous-variable quantum key distribution with 1
  mbps secure key rate}.
\newblock \emph{\bibinfo{journal}{Optics express}}
  \textbf{\bibinfo{volume}{23}}, \bibinfo{pages}{17511--17519}
  (\bibinfo{year}{2015}).

\bibitem{Kumar2015}
\bibinfo{author}{Kumar, R.}, \bibinfo{author}{Qin, H.} \&
  \bibinfo{author}{All\'{e}aume, R.}
\newblock \bibinfo{title}{Coexistence of continuous variable qkd with intense
  dwdm classical channels}.
\newblock \emph{\bibinfo{journal}{New Journal of Physics}}
  \textbf{\bibinfo{volume}{17}}, \bibinfo{pages}{043027}
  (\bibinfo{year}{2015}).

\bibitem{Huang2016}
\bibinfo{author}{Huang, D.} \emph{et~al.}
\newblock \bibinfo{title}{Field demonstration of a continuous-variable quantum
  key distribution network}.
\newblock \emph{\bibinfo{journal}{Optics Letters}}
  \textbf{\bibinfo{volume}{41}}, \bibinfo{pages}{3511--3514}
  (\bibinfo{year}{2016}).

\bibitem{Karinou2017}
\bibinfo{author}{Karinou, F.} \emph{et~al.}
\newblock \bibinfo{title}{Experimental evaluation of the impairments on a qkd
  system in a 20-channel wdm co-existence scheme}.
\newblock In \emph{\bibinfo{booktitle}{2017 IEEE Photonics Society Summer
  Topical Meeting Series (SUM)}}, \bibinfo{pages}{145--146}
  (\bibinfo{publisher}{IEEE}, \bibinfo{year}{2017}).

\bibitem{Karinou2018}
\bibinfo{author}{Karinou, F.} \emph{et~al.}
\newblock \bibinfo{title}{Toward the integration of cv quantum key distribution
  in deployed optical networks}.
\newblock \emph{\bibinfo{journal}{IEEE Photonics Technology Letters}}
  \textbf{\bibinfo{volume}{30}}, \bibinfo{pages}{650--653}
  (\bibinfo{year}{2018}).

\bibitem{Eriksson2018}
\bibinfo{author}{Eriksson, T.~A.} \emph{et~al.}
\newblock \bibinfo{title}{Coexistence of continuous variable quantum key
  distribution and 7×12.5 gbit/s classical channels}.
\newblock In \emph{\bibinfo{booktitle}{2018 IEEE Photonics Society Summer
  Topical Meeting Series (SUM)}}, \bibinfo{pages}{71--72}
  (\bibinfo{publisher}{IEEE}, \bibinfo{year}{2018}).

\bibitem{Eriksson2019}
\bibinfo{author}{Eriksson, T.~A.} \emph{et~al.}
\newblock \bibinfo{title}{Wavelength division multiplexing of continuous
  variable quantum key distribution and 18.3 tbit/s data channels}.
\newblock \emph{\bibinfo{journal}{Communications Physics}}
  \textbf{\bibinfo{volume}{2}}, \bibinfo{pages}{1--8} (\bibinfo{year}{2019}).

\bibitem{Grosshans2003}
\bibinfo{author}{Grosshans, F.} \emph{et~al.}
\newblock \bibinfo{title}{Quantum key distribution using gaussian-modulated
  coherent states}.
\newblock \emph{\bibinfo{journal}{Nature}} \textbf{\bibinfo{volume}{421}},
  \bibinfo{pages}{238--241} (\bibinfo{year}{2003}).

\bibitem{hetero04}
\bibinfo{author}{Weedbrook, C.} \emph{et~al.}
\newblock \bibinfo{title}{Quantum cryptography without switching}.
\newblock \emph{\bibinfo{journal}{Physical Review Letters}}
  \textbf{\bibinfo{volume}{93}}, \bibinfo{pages}{170504}
  (\bibinfo{year}{2004}).

\bibitem{CV_protocol_review}
\bibinfo{author}{Diamanti, E.} \& \bibinfo{author}{Leverrier, A.}
\newblock \bibinfo{title}{Distributing secret keys with quantum continuous
  variables: principle, security and implementations}.
\newblock \emph{\bibinfo{journal}{Entropy}} \textbf{\bibinfo{volume}{17}},
  \bibinfo{pages}{6072--6092} (\bibinfo{year}{2015}).

\bibitem{CV_gaussian_optimality1}
\bibinfo{author}{Navascu\'{e}s, M.}, \bibinfo{author}{Grosshans, F.} \&
  \bibinfo{author}{Acin, A.}
\newblock \bibinfo{title}{Optimality of gaussian attacks in continuous-variable
  quantum cryptography}.
\newblock \emph{\bibinfo{journal}{Physical Review Letters}}
  \textbf{\bibinfo{volume}{97}}, \bibinfo{pages}{190502}
  (\bibinfo{year}{2006}).

\bibitem{CV_gaussian_optimality2}
\bibinfo{author}{Garc\'{i}a-Patr\'{o}n, R.} \& \bibinfo{author}{Cerf, N.~J.}
\newblock \bibinfo{title}{Unconditional optimality of gaussian attacks against
  continuous-variable quantum key distribution}.
\newblock \emph{\bibinfo{journal}{Physical Review Letters}}
  \textbf{\bibinfo{volume}{97}}, \bibinfo{pages}{190503}
  (\bibinfo{year}{2006}).

\bibitem{Gaussian_unitary}
\bibinfo{author}{Leverrier, A.}
\newblock \bibinfo{title}{Security of continuous-variable quantum key
  distribution via a gaussian de finetti reduction}.
\newblock \emph{\bibinfo{journal}{Physical Review Letters}}
  \textbf{\bibinfo{volume}{118}}, \bibinfo{pages}{200501}
  (\bibinfo{year}{2017}).

\bibitem{Lev}
\bibinfo{author}{Jouguet, P.}, \bibinfo{author}{Kunz-Jacques, S.},
  \bibinfo{author}{Diamanti, E.} \& \bibinfo{author}{Leverrier, A.}
\newblock \bibinfo{title}{Analysis of imperfections in practical
  continuous-variable quantum key distribution}.
\newblock \emph{\bibinfo{journal}{Physical Review A}}
  \textbf{\bibinfo{volume}{86}}, \bibinfo{pages}{032309}
  (\bibinfo{year}{2012}).

\bibitem{Kaur2019}
\bibinfo{author}{Kaur, E.}, \bibinfo{author}{Guha, S.} \&
  \bibinfo{author}{Wilde, M.~M.}
\newblock \bibinfo{title}{Asymptotic security of discrete-modulation protocols
  for continuous-variable quantum key distribution} (\bibinfo{year}{2019}).
\newblock \eprint{1901.10099}.

\bibitem{Silberhorn2002}
\bibinfo{author}{Silberhorn, C.}, \bibinfo{author}{Ralph, T.~C.},
  \bibinfo{author}{L^^c3^^bctkenhaus, N.} \& \bibinfo{author}{Leuchs, G.}
\newblock \bibinfo{title}{Continuous variable quantum cryptography: Beating the
  3 db loss limit}.
\newblock \emph{\bibinfo{journal}{Physical Review Letters}}
  \textbf{\bibinfo{volume}{89}}, \bibinfo{pages}{167901}
  (\bibinfo{year}{2002}).

\bibitem{Hirano2003}
\bibinfo{author}{Hirano, T.}, \bibinfo{author}{Yamanaka, H.},
  \bibinfo{author}{Ashikaga, M.}, \bibinfo{author}{Konishi, T.} \&
  \bibinfo{author}{Namiki, R.}
\newblock \bibinfo{title}{Quantum cryptography using pulsed homodyne
  detection}.
\newblock \emph{\bibinfo{journal}{Physical Review A}}
  \textbf{\bibinfo{volume}{68}}, \bibinfo{pages}{042331}
  (\bibinfo{year}{2003}).

\bibitem{Leverrier2009}
\bibinfo{author}{Leverrier, A.} \& \bibinfo{author}{Grangier, P.}
\newblock \bibinfo{title}{Unconditional security proof of long-distance
  continuous-variable quantum key distribution with discrete modulation}.
\newblock \emph{\bibinfo{journal}{Physical Review Letters}}
  \textbf{\bibinfo{volume}{102}}, \bibinfo{pages}{180504}
  (\bibinfo{year}{2009}).

\bibitem{two_state_Lutken}
\bibinfo{author}{Zhao, Y.-B.}, \bibinfo{author}{Heid, M.},
  \bibinfo{author}{Rigas, J.} \& \bibinfo{author}{L^^c3^^bctkenhaus, N.}
\newblock \bibinfo{title}{Asymptotic security of binary modulated
  continuous-variable quantum key distribution under collective attacks}.
\newblock \emph{\bibinfo{journal}{Physical Review A}}
  \textbf{\bibinfo{volume}{79}}, \bibinfo{pages}{012307}
  (\bibinfo{year}{2009}).

\bibitem{three_state}
\bibinfo{author}{Br\'{a}dler, K.} \& \bibinfo{author}{Weedbrook, C.}
\newblock \bibinfo{title}{Security proof of continuous-variable quantum key
  distribution using three coherent states}.
\newblock \emph{\bibinfo{journal}{Physical Review A}}
  \textbf{\bibinfo{volume}{97}}, \bibinfo{pages}{022310}
  (\bibinfo{year}{2018}).

\bibitem{four_state_Lutken}
\bibinfo{author}{Lin, J.}, \bibinfo{author}{Upadhyaya, T.} \&
  \bibinfo{author}{L^^c3^^bctkenhaus, N.}
\newblock \bibinfo{title}{Asymptotic security analysis of discrete-modulated
  continuous-variable quantum key distribution}.
\newblock \emph{\bibinfo{journal}{Physical Review X}}
  \textbf{\bibinfo{volume}{9}}, \bibinfo{pages}{041064} (\bibinfo{year}{2019}).

\bibitem{four_state_Leverrier}
\bibinfo{author}{Ghorai, S.}, \bibinfo{author}{Grangier, P.},
  \bibinfo{author}{Diamanti, E.} \& \bibinfo{author}{Leverrier, A.}
\newblock \bibinfo{title}{Asymptotic security of continuous-variable quantum
  key distribution with a discrete modulation}.
\newblock \emph{\bibinfo{journal}{Physical Review X}}
  \textbf{\bibinfo{volume}{9}}, \bibinfo{pages}{021059} (\bibinfo{year}{2019}).

\bibitem{Papanastasiou2019}
\bibinfo{author}{Papanastasiou, P.} \& \bibinfo{author}{Pirandola, S.}
\newblock \bibinfo{title}{Continuous-variable quantum cryptography with
  discrete alphabets: Composable security under collective gaussian attacks}
  (\bibinfo{year}{2019}).
\newblock \eprint{1912.11418}.

\bibitem{Tamaki2003}
\bibinfo{author}{Tamaki, K.}, \bibinfo{author}{Koashi, M.} \&
  \bibinfo{author}{Imoto, N.}
\newblock \bibinfo{title}{Unconditionally secure key distribution based on two
  nonorthogonal states}.
\newblock \emph{\bibinfo{journal}{Physical Review Letters}}
  \textbf{\bibinfo{volume}{90}}, \bibinfo{pages}{167904}
  (\bibinfo{year}{2003}).

\bibitem{Koashi2004}
\bibinfo{author}{Koashi, M.}
\newblock \bibinfo{title}{Unconditional security of coherent-state quantum key
  distribution with a strong phase-reference pulse}.
\newblock \emph{\bibinfo{journal}{Physical Review Letters}}
  \textbf{\bibinfo{volume}{93}}, \bibinfo{pages}{120501}
  (\bibinfo{year}{2004}).

\bibitem{Shor_Preskill}
\bibinfo{author}{Shor, P.~W.} \& \bibinfo{author}{Preskill, J.}
\newblock \bibinfo{title}{Simple proof of security of the bb84 quantum key
  distribution protocol}.
\newblock \emph{\bibinfo{journal}{Physical Review Letters}}
  \textbf{\bibinfo{volume}{85}}, \bibinfo{pages}{441} (\bibinfo{year}{2000}).

\bibitem{Lo1999}
\bibinfo{author}{Lo, H.~K.} \& \bibinfo{author}{Chau, H.~F.}
\newblock \bibinfo{title}{Unconditional security of quantum key distribution
  over arbitrarily long distances}.
\newblock \emph{\bibinfo{journal}{Science}} \textbf{\bibinfo{volume}{283}},
  \bibinfo{pages}{2050--2056} (\bibinfo{year}{1999}).

\bibitem{fidelity_prev}
\bibinfo{author}{Chabaud, U.}, \bibinfo{author}{Douce, T.},
  \bibinfo{author}{Grosshans, F.}, \bibinfo{author}{Kashefi, E.} \&
  \bibinfo{author}{Markham, D.}
\newblock \bibinfo{title}{Building trust for continuous variable quantum
  states} (\bibinfo{year}{2019}).
\newblock \bibinfo{note}{1905.12700}.

\bibitem{univhash}
\bibinfo{author}{Carter, J.~L.} \& \bibinfo{author}{Wegman, M.~N.}
\newblock \bibinfo{title}{Universal classes of hash functions}.
\newblock \emph{\bibinfo{journal}{Journal of computer and system sciences}}
  \textbf{\bibinfo{volume}{18}}, \bibinfo{pages}{143--154}
  (\bibinfo{year}{1979}).

\bibitem{complementarity}
\bibinfo{author}{Koashi, M.}
\newblock \bibinfo{title}{Simple security proof of quantum key distribution
  based on complementarity}.
\newblock \emph{\bibinfo{journal}{New Journal of Physics}}
  \textbf{\bibinfo{volume}{11}}, \bibinfo{pages}{045018}
  (\bibinfo{year}{2009}).

\bibitem{Hayashi_Tsurumaru}
\bibinfo{author}{Hayashi, M.} \& \bibinfo{author}{Tsurumaru, T.}
\newblock \bibinfo{title}{Concise and tight security analysis of the
  bennett^^e2^^80^^93brassard 1984 protocol with finite key lengths}.
\newblock \emph{\bibinfo{journal}{New Journal of Physics}}
  \textbf{\bibinfo{volume}{14}}, \bibinfo{pages}{093014}
  (\bibinfo{year}{2012}).

\bibitem{Azuma_ineq}
\bibinfo{author}{Azuma, K.}
\newblock \bibinfo{title}{Weighted sums of certain dependent random variables}.
\newblock \emph{\bibinfo{journal}{Tohoku Mathematical Journal, Second Series}}
  \textbf{\bibinfo{volume}{19}}, \bibinfo{pages}{357--367}
  (\bibinfo{year}{1967}).

\bibitem{cvxpy}
\bibinfo{author}{Diamond, S.} \& \bibinfo{author}{Boyd, S.}
\newblock \bibinfo{title}{Cvxpy: A python-embedded modeling language for convex
  optimization}.
\newblock \emph{\bibinfo{journal}{Journal of Machine Learning Research}}
  \textbf{\bibinfo{volume}{17}}, \bibinfo{pages}{1--5} (\bibinfo{year}{2016}).

\bibitem{cvxpy_rewriting}
\bibinfo{author}{Agrawal, A.}, \bibinfo{author}{Verschueren, R.},
  \bibinfo{author}{Diamond, S.} \& \bibinfo{author}{Boyd, S.}
\newblock \bibinfo{title}{A rewriting system for convex optimization problems}.
\newblock \emph{\bibinfo{journal}{Journal of Control and Decision}}
  \textbf{\bibinfo{volume}{5}}, \bibinfo{pages}{42--60} (\bibinfo{year}{2018}).

\bibitem{Devetak_Winter}
\bibinfo{author}{Devetak, I.} \& \bibinfo{author}{Winter, A.}
\newblock \bibinfo{title}{Distillation of secret key and entanglement from
  quantum states}.
\newblock \emph{\bibinfo{journal}{Proceedings of the Royal Society A:
  Mathematical, Physical and engineering sciences}}
  \textbf{\bibinfo{volume}{461}}, \bibinfo{pages}{207--235}
  (\bibinfo{year}{2005}).

\bibitem{laguerre}
\bibinfo{author}{Abramowitz, M.} \& \bibinfo{author}{Stegun, I.~A.}
\newblock \emph{\bibinfo{title}{Handbook of mathematical functions with
  formulas, graphs, and mathematical tables}}, vol.~\bibinfo{volume}{55}
  (\bibinfo{publisher}{US Government printing office}, \bibinfo{year}{1948}).

\bibitem{composability}
\bibinfo{author}{M\"{u}ller-Quade, J.} \& \bibinfo{author}{Renner, R.}
\newblock \bibinfo{title}{Composability in quantum cryptography}.
\newblock \emph{\bibinfo{journal}{New Journal of Physics}}
  \textbf{\bibinfo{volume}{11}}, \bibinfo{pages}{085006}
  (\bibinfo{year}{2009}).

\bibitem{refined_Azuma_ineq1}
\bibinfo{author}{Raginsky, M.} \& \bibinfo{author}{Sason, I.}
\newblock \bibinfo{title}{Concentration of measure inequalities in information
  theory, communications, and coding}.
\newblock \emph{\bibinfo{journal}{Foundations and Trends^^c3^^82^^c2^^ae in
  Communications and Information Theory}} \textbf{\bibinfo{volume}{10}},
  \bibinfo{pages}{1--246} (\bibinfo{year}{2013}).
\newblock \urlprefix\url{http://dx.doi.org/10.1561/0100000064}.

\bibitem{refined_Azuma_ineq2}
\bibinfo{author}{McDiarmid, C.}
\newblock \emph{\bibinfo{title}{Concentration}}, \bibinfo{pages}{195--248}.
\newblock Probabilistic methods for algorithmic discrete mathematics
  (\bibinfo{publisher}{Springer}, \bibinfo{year}{1998}).

\end{thebibliography}

\medskip

\noindent{\bf \large Acknowledgements}\\
This work was supported by Cross-ministerial Strategic Innovation Promotion Program (SIP) (Council for Science, Technology and Innovation (CSTI));
ImPACT Program (CSTI);
CREST (Japan Science and Technology Agency) JPMJCR1671;
JSPS KAKENHI Grant Number JP18K13469.

\end{document}